\documentclass[%
 aip,
 amsmath,amssymb,
 reprint,%
]{revtex4-1}

\usepackage{graphicx}
\usepackage{dcolumn}
\usepackage{bm}
\usepackage{color}
 \usepackage[version=3]{mhchem}
\usepackage{setspace}
\usepackage{threeparttable}

\begin{document}

\preprint{AIP/123-QED}

\title[]{Assessment of Charge-Transfer Excitations in Organic Dyes obtained from TD-srDFT Based on Long-Range MP2 and MCSCF
Wave Functions \\
}

\author{Erik Donovan Hedeg{\aa}rd}
 \email{edh@sdu.dk}
\affiliation{Department of Physics, Chemistry and Pharmacy, University of Southern Denmark, Odense, Denmark}
\author{Frank Heiden}%
\affiliation{Department of Physics, Chemistry and Pharmacy, University of Southern Denmark, Odense, Denmark}
\author{Stefan Knecht}
\affiliation{Department of Physics, Chemistry and Pharmacy, University of Southern Denmark, Odense, Denmark}
\affiliation{Laboratory of Physical Chemistry, ETH Z{\"u}rich, Wolfgang-Pauli-Str. 10, CH-8093 Z{\"u}rich, Switzerland}
\author{Emmanuel Fromager}
\affiliation{Laboratoire de Chimie Quantique, 
Institut de Chimie, CNRS et Universit{\'e} de Strasbourg,
4 rue Blaise Pascal, 67000 Strasbourg, France
}
\author{Hans J{\o}rgen Aagaard Jensen}
 \email{hjj@sdu.dk}
\affiliation{Department of Physics, Chemistry and Pharmacy, University of Southern Denmark, Odense, Denmark}

\date{\today}

\begin{abstract}
  Charge transfer excitations can be described within Time-Dependent
  Density Functional Theory (TD-DFT),
  not only by means of long-range corrected exchange functionals but also 
with a combination of wave function theory and TD-DFT based on range
separation. The latter approach enables a rigorous formulation of
multi-determinantal TD-DFT schemes where excitation classes, which are absent in conventional TD-DFT spectra (like for example double
excitations), can be
addressed. 
This paper investigates the combination of both the long-range {\it
Multi-Configuration Self-Consistent Field} (MCSCF) and {\it Second Order
Polarization Propagator Approximation} (SOPPA) \emph{ans\"atze}\ with a
short-range DFT (srDFT) description. 
We find that the combinations of SOPPA or MCSCF with TD-DFT yield better
results than could be expected from the pure wave function schemes. For
the Time-Dependent MCSCF short-range DFT \textit{ansatz} (TD-MC-srDFT) 
excitation energies calculated over a larger benchmark set of molecules with predominantly single reference character yield good agreement with their reference values, and are in general comparable 
to the long-range corrected functional CAM-B3LYP. The SOPPA-srDFT scheme is tested for a subset of molecules used for benchmarking 
TD-MC-srDFT and performs slightly better against the reference data for this small subset. 
Beyond the proof-of-principle calculations comprising the first part of this contribution, we additionally studied the low-lying singlet excited states ($S_1$\ and $S_2$) 
of the retinal chromophore.  
The chromophore displays multireference character in the ground state and both excited states 
exhibit considerable double excitation character, which in turn cannot
be described within standard TD-DFT, due to the adiabatic approximation. 
However, a TD-MC-srDFT approach can account for the multireference character, and excitation energies are obtained with 
accuracy comparable to CASPT2, although using a much smaller active space.

\end{abstract}

\pacs{Valid PACS appear here}
\maketitle

%

\section{\label{intro} Introduction}

The energy absorption in the electronic excitations wave number regime is the basis for numerous industrial applications such as
dye-sensitized solar cells and artificial photo synthesis\cite{graetzel1981,gust2009}.
  In these areas, theoretical chemistry holds a great potential in the prediction and fine 
tuning of new molecular building blocks for novel materials. 
To link experiment with theory, it is of utmost importance to have methods at hand which can predict accurate electron excitation energies for several 
excitation classes within a given molecule. 

The success of Time-Dependent Density Functional Theory (TD-DFT) in this
area\cite{runge1984,marques2004,Casida_tddft_review_2012} relies on its 
accuracy and computational efficiency for excitations occurring between
orbitals within a functional group or between functional groups in close
proximity to each other. Such excitations 
are typically referred to as the class of \textit{local excitations}. Other examples of excitation classes are Rydberg and Charge Transfer (CT) excitations. 
For these types of excitations TD-DFT is known to exhibit shortcomings: 
Rydberg excitations are notoriously underestimated due to the wrong 
asymptotic behavior of most Generalized Gradient Approximated (GGA) DFT functionals.
Also CT excitations are often severely underestimated by TD-DFT\cite{tozer1999,liao2003,fabiano2006,perpete2006}. 
A detailed study by Dreuw and Head-Gordon in 2004~\cite{dreuw2004} showed that
regular exchange-correlation density-functionals display a wrong
behavior of CT states with respect to the distance between the separated
charges and it was argued that this feature is inherently caused by  
a self-interaction error arising through the electron transfer in the CT
state, ultimately leading to underestimated CT
excitation energies.  Errors in Rydberg excitations can to some degree
be remedied by asymptotically
 corrected functionals, and new functionals have also been proposed to meet the challenges
for CT excitations. Examples include the range-separated hybrid
functionals\cite{vydrov2006,rohrdanz2009,baer2010} such as
CAM-B3LYP\cite{yanai2004} that rely on the separation of the
two-electron repulsion $1/r_{12}$ into long- and short-range parts for the
calculation of the exchange energy. 
The above-mentioned developments have been driven by modifying exchange functionals to meet the requirements within the various excitation classes.
 However, TD-DFT still exhibits some fundamental flaws, which are difficult
 to overcome in the conventional single-configuration Kohn-Sham (KS) framework.
For instance, for compounds with significant amount of multireference character and/or with excitations which have a high degree of double excitation character, TD-DFT will generally fail
within the adiabatic approximation,\cite{maitra2004,neugebauer2004,elliot2011,burke2012} which is the standard approach 
in most quantum chemistry codes today.
Suitable methods such as the {\it Multi-Configuration Self-Consistent
Field} (MCSCF) approach suffer on the other hand from a neglect of large parts of the local
dynamical correlation which has to be recovered in a subsequent step. Popular approaches to achieve this goal are mainly based on multireference perturbation theory and we shall 
here mention {\it second-order Complete Active Space perturbation
theory} (CASPT2)~\cite{andersson1990,andersson1992} and {\it second-order N-electron
valence state perturbation theory} (NEVPT2)~\cite{angeli2001}. 
Nevertheless, these methods are already computationally expensive  
whereas more accurate multireference Configuration Interaction (MRCI) or Coupled Cluster (MRCC) schemes 
are even more restricted due to their steep exponential scaling with system size. 

An appealing alternative consists thus in coupling wave function theory 
(WFT) with DFT where the latter accounts for the major part of the dynamical correlation.
There have been several suggestions for how such a scheme could be devised, e.g.~the DFT/MRCI method developed by Grimme and co-workers\cite{grimme1999,marian2008}.
 We focus in this work on the so-called long-range WFT / short-range DFT
 (WFT-srDFT) approach for which  
long-range Hartree-Fock (HF) and post-HF approximations have been
developed in recent years by various research groups. Currently, the methods 
HF-srDFT~\cite{angyan}, {\it Second-order M\o ller-Plesset} srDFT
(MP2-srDFT)~\cite{angyan,fromager2008},
CI-srDFT~\cite{leininger1997}, CC-srDFT~\cite{goll2005},
MC-srDFT~\cite{fromager2007,fromager2009} and 
NEVPT2-srDFT~\cite{nevpt2srdft} have been presented. 
The extension to the time-dependent linear response regime has been
 explored initially by Pernal~\cite{JCP12_Pernal_tddmft-srdft} who described long-range
correlation effects within one-electron reduced density-matrix
theory. Very recently time-dependent versions of 
HF-srDFT~\cite{fromager2013,td-hf-srdft_open_shell_Elisa}
and MC-srDFT schemes~\cite{fromager2013} have been implemented. We denoted these
time-dependent
methods TD-HF-srDFT and TD-MC-srDFT, respectively.

In this work we investigate the performance of the TD-MC-srDFT method for calculation of local and CT
excitations in organic dyes. Comparison is made with standard TD-DFT
results, obtained with the regular hybrid B3LYP and the long-range
corrected hybrid CAM-B3LYP functionals. We further introduce the SOPPA-srDFT scheme, whose formulation is
based on a long-range MP2 expansion of the TD-MC-srDFT~\cite{fromager2013} linear response
equations. The SOPPA-srDFT method is tested against a subset of the molecules used to benchmark TD-MC-srDFT. 
As a final application, we investigate the performance of the TD-MC-srDFT method for the retinal chromophore, 
for which CASPT2 has been the standard method of choice for long\cite{ferre2003}.
The theory behind the TD-MC-srDFT and SOPPA-srDFT methods are summarized in the following section (Section \ref{theory}), while computational details for 
the benchmark set of molecules and the retinal chromophore 
are given in Section \ref{Compdetails}. All results are described in
Section \ref{results}, and conclusions are given in the final section (Section \ref{conclusion}).

\section{\label{theory}Theory}

\subsection{Range-separated density-functional theory}\label{subsec:srdft}

The multi-determinantal extensions of TD-DFT considered in this work 
rely on the range separation of the regular two-electron
repulsion~\cite{savinbook} 
\begin{equation}
\vert\mathbf{r} - \mathbf{r}'\vert^{-1 } =  w^{\rm lr,\mu}_{\rm
ee}(\vert\mathbf{r} - \mathbf{r}'\vert)+w^{\rm sr,\mu}_{\rm
ee}(\vert\mathbf{r} - \mathbf{r}'\vert) , 
\label{sr-DFTCoulpart}
\end{equation}
where the long-range interaction based on the error function is used,
\begin{eqnarray}
w^{\rm lr,\mu}_{\rm ee}(\vert\mathbf{r} - \mathbf{r}'\vert)&=&\frac{\rm
erf \bigl(\mu\, \vert\mathbf{r} - \mathbf{r}'\vert \bigr)
}{\vert\mathbf{r} - \mathbf{r}'\vert}, 
\end{eqnarray}
and $\mu$ is the parameter that controls the range separation. The exact
ground-state energy of an electronic system can then in principle be obtained
variationally as follows   
\begin{align}
 E = \min_{\Psi} & \Big\{ \langle\Psi\vert\hat{T} 
+ \hat{W}^{\text{lr},\mu}_{\rm
ee}\vert\Psi\rangle  + E_{\rm Hxc}^{\text{sr},\mu}[\rho_{\Psi}]   \notag \\
&  
+  \int\text{d}\mathbf{r} \,v_{\rm
ne}(\mathbf{r})\rho_{\Psi}(\mathbf{r}) \Big\}, 
\label{sr-DFTenergy}
\end{align}
where $\hat{T}$ and $\hat{W}^{\rm lr,\mu}_{\rm ee}$ are the kinetic energy 
and long-range two-electron
interaction operators, respectively, while $v_{\rm ne}(\mathbf{r})$
denotes the local nuclear potential. The $\mu$-dependent complementary 
density-functional $E^{\rm sr,\mu}_{\rm Hxc}[\rho]$ can be split into
short-range Hartree, exchange and correlation (srHxc) contributions 
\begin{equation}
E^{\rm sr,\mu}_{\rm Hxc}[\rho] = E^{\text{sr},\mu}_{\text{H}}[\rho]  +
E^{\text{sr},\mu}_{\text{x}}[\rho]
+E^{\text{sr},\mu}_{\text{c}}[\rho]
,  
\end{equation}
where $E^{\rm sr,\mu}_{\rm H}[\rho]= \frac{1}{2}\int\int
{\rm d}{\mathbf{r}}{\rm d}{\mathbf{r'}}\rho(\mathbf{r})\rho(\mathbf{r'})w^{\rm
sr,\mu}_{\rm ee}\left(\vert {\bf r}-{\bf r'} \vert\right)$. The usual 
expression for the exact short-range exchange energy
$E^{\text{sr},\mu}_{\text{x}}[\rho]=\langle\Phi^{\rm KS}[\rho]\vert\hat{W}^{\text{sr},\mu}_{\rm
ee}\vert \Phi^{\rm KS}[\rho] \rangle$ is, as in standard DFT, based
on the KS determinant. This definition has been used by Savin and
co-workers~\cite{erferfgaufunc} for constructing
approximate short-range exchange and correlation density-functionals. 
The exact minimizing wave function $\Psi^{\mu}$ in Eq.~(\ref{sr-DFTenergy})
is the ground state  of the long-range interacting system whose density
$\rho^{\mu}(\mathbf{r})=\langle\Psi^\mu\vert\hat{\rho}(\mathbf{r})\vert\Psi^\mu\rangle$
equals the density of the physical fully-interacting system. It fulfills the self-consistent equation 
\begin{eqnarray}\begin{array}{l}\label{srdft_sc_eq}
\hat{H}^{\mu}[\rho^{\mu}]\vert\Psi^\mu\rangle=\mathcal{E}^\mu\vert\Psi^\mu\rangle,
\end{array}
\end{eqnarray}
where the density-dependent long-range Hamiltonian equals 
\begin{eqnarray}\begin{array}{l}\label{srdft_sc_eq2}
\hat{H}^{\mu}[\rho]=\hat{T} +
\hat{W}^{\text{lr},\mu}_{\rm ee} + \hat{V}_{\rm
ne}+\hat{V}_{\rm Hxc}^{\text{sr},\mu}[\rho],\\
\\
\displaystyle
 \hat{V}_{\rm Hxc}^{\text{sr},\mu}[\rho] =  \int\text{d}\mathbf{r}\,\frac{\delta
 E^{\text{sr},\mu}_{\rm Hxc}}{\delta
 \rho(\mathbf{r})}[\rho]\,\hat{\rho}(\mathbf{r}),
\end{array}
\end{eqnarray}
and $\hat{V}_{\rm
ne}=\int {\rm d}{\bf r}\,v_{\rm ne}({\bf r})\,\hat{\rho}({\bf r})$.
 Since the long-range interaction is treated explicitly, in contrast to KS-DFT, the exact
solution is multi-determinantal. 
The approximate range-separated DFT models considered in this work describe
 the long-range interaction at the HF, MP2 and MCSCF
levels. These schemes will be referred to as HF-srDFT, MP2-srDFT and
MC-srDFT, respectively. Their extensions to the time-dependent linear
response regime is
presented in the following.

\subsection{Extension to the time-dependent regime 
}\label{subsec:exacttd-srdft}

As discussed in details in Ref.~\onlinecite{fromager2013}, excited-state properties can be
described 
when extending Eqs.~(\ref{sr-DFTenergy}) and
(\ref{srdft_sc_eq}) to the time-dependent
regime. Within the short-range adiabatic
approximation~\cite{fromager2013}, the
time evolution of the auxiliary long-range interacting system is
obtained as
follows
\begin{align}\label{eq:tdlreq}
 \left(\hat{T} + \hat{W}^{\text{lr},\mu}_{\rm ee} + \hat{V}(t)\right. &
 + \left.\hat{V}^{\text{sr},\mu}_{\text{Hxc}}[\tilde{\rho}^{\mu}(\mathbf{r},t)]
 -{\rm i}\frac{\partial}{\partial t} \right) 
\vert\tilde{\Psi}^{\mu}(t)\rangle \notag \\ 
& = Q^{\mu}(t) \vert\tilde{\Psi}^{\mu}(t)\rangle,
\end{align}
where
$\tilde{\rho}^{\mu}(\mathbf{r},t)=\langle\tilde{\Psi}^{\mu}(t)\vert\hat{\rho}({\bf
r})\vert\tilde{\Psi}^{\mu}(t)\rangle$ is an approximation to the exact
time-dependent density of the physical fully-interacting system and
$
  \hat{V}(t) =  \int {\rm
  d}\mathbf{r}\,v(\mathbf{r},t)\,\hat{\rho}(\mathbf{r})
$
is a local time-dependent potential operator. If the latter is periodic
of period $T$, Eq.~(\ref{eq:tdlreq}) is equivalent to the variational
principle 
\begin{equation}\label{td-vp-sradia}
 \delta \mathcal{Q}^{\mu}[\tilde{\Psi}^{\mu}] = 0,
\end{equation}
which is formulated in terms of the range-separated 
wave function-dependent action integral 
\begin{eqnarray}\label{quasienerAsrdft}\begin{array} {l}
\displaystyle\mathcal{Q}^{\mu}[{{{\Psi}}}]=\int_{0}^{T}
{\rm d}t\; \frac{\langle {{\Psi}(t)}\vert \hat{T}+{\hat{W}^{\rm
lr,\mu}_{\rm ee}}+\hat{V}(t)-{\rm i}\frac{\partial}{\partial t}\vert
{{\Psi}(t)}\rangle}{\langle
{{\Psi}(t)}\vert{{\Psi}(t)} \rangle}\\
\\
\displaystyle
\hspace{1.3cm}+\int_{0}^{T} {\rm d}t\; {E^{\rm sr,\mu}_{\rm
Hxc}}[{\rho_{{\Psi}(t)}}]. \\
\end{array}
\end{eqnarray}
The linear response TD-MC-srDFT model discussed in the following
is based on the variational formulation in Eq.~(\ref{td-vp-sradia}).
\subsection{TD-MC-srDFT model}\label{subsec:td-mc-srdft}
We work in this section in the framework of Floquet
theory~\cite{christiansen1998} where the time-dependent
periodic perturbation is decomposed as follows 
\begin{eqnarray}\label{periodicpert1}\begin{array} {l}
\displaystyle{{\hat{V}}(t)=\hat{V}_{\rm ne}
  +\sum_{x}\sum^N_{k=-N}e^{-{\rm i}\omega_kt}\varepsilon_x(\omega_k)\hat{V}_x},\\
\displaystyle \omega_k=\frac{2\pi k}{T},\\
\\
\displaystyle
\hat{V}_x=\int {\rm d}\mathbf{r}\; v_x(\mathbf{r})\hat{\rho}(\mathbf{r}).
\end{array}
\end{eqnarray}
We use a MCSCF parametrization of the time-dependent wave function $\tilde{\Psi}^{\mu}(t)\rightarrow \vert\tilde{0}^{\mu}(t) \rangle$
in Eq.~(\ref{td-vp-sradia}) consisting of exponential unitary
transformations~\cite{olsen1985} 
\begin{eqnarray}\label{tdmcparametr}\begin{array} {l}
\vert\tilde{0}^{\mu}(t)\rangle=e^{{\rm i}\hat{\kappa}(t)}e^{{\rm
i}\hat{S}(t)}\vert0^{\mu}\rangle,
\end{array}
\end{eqnarray}
which are applied to the unperturbed MC-srDFT wave function $\vert 0^{\mu}\rangle$ with
\begin{eqnarray}\label{tdrotation}\begin{array} {l}
{\displaystyle
\hat{\kappa}(t)=\sum_{l,i}e^{-{\rm i}\omega_l
t}\kappa_{i}(\omega_l)\hat{q}_i^{\dagger}+e^{-{\rm i}\omega_l
t}\kappa^*_{i}(-\omega_l)\hat{q}_i,
}
\\
\\
{\displaystyle
\hat{S}(t)=\sum_{l,i}e^{-{\rm i}\omega_l
t}S_{i}(\omega_l)\hat{R}_i^{\dagger}+e^{-{\rm i}\omega_l
t}S^*_{i}(-\omega_l)\hat{R}_i.
}
\end{array}
\end{eqnarray}
The singlet excitation and state-transfer operators are defined as follows 
\begin{eqnarray}\label{qidagridag}\begin{array} {l}
{\displaystyle
\hat{q}_i^{\dagger}=\hat{E}_{pq}=\hat{a}^{\dagger}_{p\alpha}\hat{a}_{q\alpha}+\hat{a}^{\dagger}_{p\beta}\hat{a}_{q\beta};
\;\; p>q},\\
\\
\hat{R}_i^{\dagger}=\vert i \rangle\langle 0^{\mu}\vert. 
\end{array}
\end{eqnarray}
Note that the TD-HF-srDFT scheme is a particular case of
Eq.~(\ref{tdmcparametr}), where the unperturbed MC-srDFT wave function would be
replaced by the HF-srDFT determinant, and only orbital rotations would
be considered. 
Returning to the
multi-configuration case, the TD-MC-srDFT wave function in
Eq.~(\ref{tdmcparametr}) is fully determined by the
Fourier component vectors
\begin{eqnarray}\label{rspvec}\begin{array} {l}
\Lambda(\omega_l)= \begin{bmatrix}
\kappa_{i}(\omega_l)\\
S_{i}(\omega_l) \\
\kappa^*_{i}(-\omega_l)\\
S^*_{i}(-\omega_l)	  \end{bmatrix},
	  \\
\end{array}
\end{eqnarray}
for which we consider in the following the Taylor expansion through
first order:
\begin{eqnarray}\label{linearrspvec}\begin{array} {l}
$${\displaystyle{\Lambda(\omega_l)}=\sum^N_{k=-N,x}\varepsilon_x(\omega_k){\left
. \frac{\partial {\Lambda(\omega_l)}}{\partial
\varepsilon_x(\omega_k)}\right |_{\bm \varepsilon = 0}}+ \ldots}$$
\end{array}
\end{eqnarray}
Rewriting the variational condition in Eq.~(\ref{td-vp-sradia}) as
follows
\begin{eqnarray}\label{varcondquasisrdft}\begin{array} {l}
{\displaystyle\forall\;\varepsilon_x(\omega_k) \;\;\;\;\;\;\frac{\partial \mathcal{Q}^{\rm \mu}}{\partial
{\Lambda^{\dagger}(-\omega_l)
}}
=0}, 
\end{array}
\end{eqnarray}
the linear response equations are simply obtained by differentiation
with respect to the perturbation strength
$\varepsilon_x(\omega_k)$~\cite{fromager2013}: 
\begin{eqnarray}\label{linearrspsrdft}\begin{array} {l}
\displaystyle \left(\left .
\frac{\rm d}{{\rm d}\varepsilon_x(\omega_k)}\frac{\partial \mathcal{Q}^{\rm
\mu}}{\partial {\Lambda^{\dagger}(-\omega_l)
}}\right)\right |_{\bm \varepsilon = 0}=0, 
\end{array}
\end{eqnarray}
which leads, according to Eq.~(\ref{quasienerAsrdft}) and
Refs.~\onlinecite{fromager2013,jcp02_Trond_tdhf}, to 
\begin{eqnarray}\label{derivaQlr_plus_sr}\begin{array} {l}
\displaystyle \left(\left .
\frac{\rm d}{{\rm d}\varepsilon_x(\omega_k)}\frac{\partial {\mathcal{Q}^{\rm
\mu}}}{\partial {\Lambda^{\dagger}(-\omega_l)}}\right)\right |_{\bm
\varepsilon = 0}=
\frac{\rm d}{{\rm d}\varepsilon_x(\omega_k)}\frac{\partial }{\partial
{\Lambda^{\dagger}(-\omega_l)}}
\\
\\
\Bigg(\displaystyle
\displaystyle
\frac{T}{2}\sum_{m,n}\delta(\omega_m+\omega_n)
\\
\displaystyle
\hspace{0.4cm}\times \Lambda^{\dagger}(-\omega_m)\Big[E^{[2]\mu}_0+{K_{\rm Hxc}^{\rm
sr,\mu}}+\omega_m S^{[2]\mu}\Big]\Lambda(\omega_n)\\
\\
\displaystyle
\hspace{0.2cm}+\frac{T}{2}\sum_{m}\sum_{y}\sum^N_{p=-N}\delta(\omega_m+\omega_p)\varepsilon_y(\omega_p)
\\
\times \Big[
{\rm i}V_y^{[1]\mu\dagger}\Lambda(\omega_m)-{\rm
i}\Lambda^{\dagger}(-\omega_m)V_y^{[1]\mu}
\Big]
\left.\Bigg)\right |_{\bm \varepsilon = 0}=0. 
\end{array}
\end{eqnarray}
Several matrices and vectors have been introduced in Eq.~\eqref{derivaQlr_plus_sr}. First are the long-range Hessian, $E_0^{[2]\mu}$, the srHxc kernel, 
$K^{\text{sr},\mu}_{\text{Hxc}}$, and $\mu$-dependent metric $S^{[2]\mu}$. 
These matrices will along with the property gradient vector, $V^{[1]\mu}_{y}$ be described in more detail below:
The long-range Hessian ($E_0^{[2]\mu}$) is obtained from the auxiliary
Hamiltonian $\hat{H}_0^{\mu}=\hat{H}^{\mu}[\rho_0^\mu]$, that is calculated for the 
unperturbed MC-srDFT density
$\rho_0^\mu(\mathbf{r})=\langle0^\mu\vert\hat{\rho}(\mathbf{r})\vert0^\mu\rangle$, as follows 
\begin{eqnarray}\label{lrHessianmu}\begin{array} {l}
E_0^{[2]\mu}= \begin{bmatrix}
A^\mu&B^\mu\\
B^{\mu*}&A^{\mu*} \\
\end{bmatrix},
\end{array}
\end{eqnarray}
\begin{eqnarray}\label{ABmu}\begin{array} {l}
\hspace{-0.2cm}
A^{\mu}= \begin{bmatrix}
\langle 0^\mu \vert
[\hat{q}_{i},[\hat{H}_0^{\mu},\hat{q}^{\dagger}_{j}]]\vert
0^\mu\rangle 
& 
\langle 0^\mu \vert
[[\hat{q}_{i},\hat{H}_0^{\mu}],\hat{R}^{\dagger}_{j}]\vert
0^\mu\rangle 
\\
\langle 0^\mu \vert
[\hat{R}_{i},[\hat{H}_0^{\mu},\hat{q}^{\dagger}_{j}]]\vert
0^\mu\rangle 
&
\langle 0^\mu \vert
[\hat{R}_{i},[\hat{H}_0^{\mu},\hat{R}^{\dagger}_{j}]]\vert
0^\mu\rangle 
\\
\end{bmatrix}
,\\
\\
\hspace{-0.2cm}
B^{\mu}= \begin{bmatrix}
\langle 0^\mu\vert
[\hat{q}_{i},[\hat{H}_0^{\mu},\hat{q}_{j}]]\vert 0^\mu\rangle 
& 
\langle 0^\mu\vert [[\hat{q}_{i},\hat{H}_0^{\mu}],\hat{R}_{j}]\vert
0^\mu\rangle 
\\
\langle 0^\mu \vert
[\hat{R}_{i},[\hat{H}_0^{\mu},\hat{q}_{j}]]\vert 0^\mu\rangle 
&
\langle 0^\mu \vert
[\hat{R}_{i},[\hat{H}_0^{\mu},\hat{R}_{j}]]\vert0^\mu\rangle 
\\
\end{bmatrix}
,
\end{array}
\end{eqnarray}
and the $\mu$-dependent metric equals 
\begin{eqnarray}\label{metricmu}\begin{array} {l}
S^{[2]\mu}= \begin{bmatrix}
\Sigma^\mu&\Delta^\mu\\
-\Delta^{\mu*}&-\Sigma^{\mu*}\\
\end{bmatrix},
\end{array}
\end{eqnarray}
\begin{eqnarray}\label{SDmu}\begin{array} {l}
\Sigma^{\mu}= \begin{bmatrix}
\langle 0^\mu \vert [\hat{q}_{i}
,\hat{q}^{\dagger}_{j}]\vert 0^\mu\rangle 
& 
\langle 0^\mu \vert [\hat{q}_{i}
,\hat{R}^{\dagger}_{j}]\vert 0^\mu\rangle 
\\
\langle 0^\mu \vert [\hat{R}_{i}
,\hat{q}^{\dagger}_{j}]\vert 0^\mu\rangle 
&
\langle 0^\mu \vert [\hat{R}_{i}
,\hat{R}^{\dagger}_{j}]\vert 0^\mu\rangle 
\\
\end{bmatrix},
\\
\\
\Delta^{\mu}= \begin{bmatrix}
\langle 0^\mu \vert [\hat{q}_{i}
,\hat{q}_{j}]\vert 0^\mu\rangle 
& 
\langle 0^\mu \vert [\hat{q}_{i}
,\hat{R}_{j}]\vert 0^\mu\rangle 
\\
\langle 0^\mu \vert [\hat{R}_{i}
,\hat{q}_{j}]\vert 0^\mu\rangle 
&
\langle 0^\mu \vert [\hat{R}_{i}
,\hat{R}_{j}]\vert 0^\mu\rangle 
\\
\end{bmatrix}
.
\end{array}
\end{eqnarray}
The srHxc kernel
contribution in Eq.~(\ref{derivaQlr_plus_sr}) is calculated for the unperturbed MC-srDFT density, 
\begin{eqnarray}\label{srkernelHessian}\begin{array} {l}
\displaystyle
K_{\rm Hxc}^{\rm sr,\mu}=
\int\int {\rm d}\mathbf{r}{\rm d}\mathbf{r'}\; K_{\rm Hxc}^{\rm
sr,\mu}[\rho_0^\mu](\mathbf{r},\mathbf{r'})\rho^{[1]\mu}(\mathbf{r})\rho^{[1]\mu\dagger}(\mathbf{r'}),
\\
\\
\displaystyle
K_{\rm Hxc}^{\rm sr,\mu}[\rho](\mathbf{r},\mathbf{r'})=\frac{\delta^2 E_{\rm
Hxc}^{\rm sr,\mu}}{\delta \rho(\mathbf{r})\delta
\rho(\mathbf{r'})}[\rho],
\end{array}
\end{eqnarray}
and (as seen in Eq.~\ref{srkernelHessian} above), expressed in terms of
the gradient density vector 
\begin{eqnarray}\label{gradientdensvector}\begin{array} {l}
\rho^{[1]\mu}(\mathbf{r})= \begin{bmatrix}
\langle0^{\mu}\vert[\hat{q}_i,\hat{\rho}(\mathbf{r})]\vert0^{\mu}\rangle\\
\langle0^{\mu}\vert[\hat{R}_i,\hat{\rho}(\mathbf{r})]\vert0^{\mu}\rangle\\
\langle0^{\mu}\vert[\hat{q}^{\dagger}_i,\hat{\rho}(\mathbf{r})]\vert0^{\mu}\rangle\\
\langle0^{\mu}\vert[\hat{R}^{\dagger}_i,\hat{\rho}(\mathbf{r})]\vert0^{\mu}\rangle\\
\end{bmatrix}
.
\end{array}
\end{eqnarray}
Finally, the gradient property vector equals 
\begin{equation}
V_y^{[1]\mu} = \int {\rm d}\mathbf{r}\;
v_y(\mathbf{r})\rho^{[1]\mu}(\mathbf{r}). \label{prop-gradient}
\end{equation} 
The linear response equations (Eq.~\ref{derivaQlr_plus_sr}) can now
be rewritten in a compact form as follows
\begin{eqnarray}\label{lr_srdft_eq}
\hspace{-0.7cm} \displaystyle
\left . \Bigg(E^{[2]\mu}+\omega_lS^{[2]\mu}\Bigg)\frac{\partial
\Lambda(-\omega_l)}{\partial \varepsilon_x(\omega_k)}\right |_{\bm \varepsilon = 0} &=&{\rm i}
V_x^{[1]\mu}\delta(\omega_k+\omega_l),
\end{eqnarray}
or, equivalently,
\begin{eqnarray}\label{lr_srdft_eq_minusomega}\begin{array} {l}
\hspace{-0.7cm} \displaystyle
\left . \Bigg(E^{[2]\mu}-\omega_lS^{[2]\mu}\Bigg)
\frac{\partial
\Lambda(\omega_l)}{\partial \varepsilon_x(\omega_k)}\right |_{\bm \varepsilon = 0} 
={\rm i}
V_x^{[1]\mu}\delta(\omega_k-\omega_l),
\end{array}
\end{eqnarray}
where the MC-srDFT Hessian is comprised of the long-range Hessian and the Hxc kernel from Eqs.~\eqref{lrHessianmu} and \eqref{srkernelHessian}
\begin{eqnarray}\label{srdftHessian}\begin{array} {l}
E^{[2]\mu}=E_0^{[2]\mu}+K_{\rm Hxc}^{\rm sr,\mu}.
\end{array}
\end{eqnarray}
Note that in Eq.~(38) of Ref.~\citenum{fromager2013} the metric that was
used is the one in Eq.~(\ref{metricmu}) multiplied by -1, as in
Ref.~\citenum{jcp02_Trond_tdhf}. This is why
the metric is multiplied by $+\omega_l$ in Eq.~(\ref{lr_srdft_eq})
instead of $-\omega_l$ as done in Ref.~\citenum{fromager2013}. 

The time-dependent expectation value of the perturbation can thus be
expanded through first order
\begin{eqnarray}\label{kubo}\begin{array}{l}
\langle\tilde{0}^{\mu}(t)\vert\hat{V}_y\vert\tilde{0}^{\mu}(t)\rangle
=
\langle{0}^{\mu}\vert\hat{V}_y\vert{0}^{\mu}\rangle
\\
\\
\displaystyle
+{\rm i}\sum_l
e^{-{\rm
i}\omega_lt}V_y^{[1]\mu\dagger}
\sum_{x}\sum^N_{k=-N}
\varepsilon_x(\omega_k)
\left.\frac{\partial
\Lambda(\omega_l)}{\partial \varepsilon_x(\omega_k)}\right |_{\bm \varepsilon = 0} 
\\
+\ldots
\\
\\
\displaystyle
=\langle{0}^{\mu}\vert\hat{V}_y\vert{0}^{\mu}\rangle
+\sum_{x}\sum^N_{k=-N}e^{-{\rm
i}\omega_kt}\varepsilon_x(\omega_k)\langle\langle\hat{V}_y,\hat{V}_x\rangle\rangle_{\omega_k}\\
+\ldots
,
\end{array}
\end{eqnarray}
where, according to Eq.~(\ref{lr_srdft_eq_minusomega}), the linear response function equals
\begin{eqnarray}\label{linear_rsp_function}\begin{array} {l}
\displaystyle
\langle\langle\hat{V}_y,\hat{V}_x\rangle\rangle_{\omega_k}=
-V_y^{[1]\mu\dagger}\Big[E^{[2]\mu}-\omega_kS^{[2]\mu}\Big]^{-1}
V_x^{[1]\mu}.
\end{array}
\end{eqnarray}
Excitation energies $\omega_I$ can then be calculated at the TD-MC-srDFT level when 
solving iteratively
\begin{eqnarray}\label{MCCasidaeq}\begin{array} {l}
\displaystyle
\Big(E^{[2]\mu}-\omega_I S^{[2]\mu}\Big)X(\omega_I)=0.\label{exc-energies}
\end{array}
\end{eqnarray}
The linear response function in Eq.~(\ref{linear_rsp_function}) can \textit{formally} be re-expressed in
the basis of the converged solutions $X(\omega_I)$ which leads
to~\cite{olsen1985}
\begin{eqnarray}\label{linear_rsp_function_diag}\begin{array} {l}
\displaystyle
\langle\langle\hat{V}_y,\hat{V}_x\rangle\rangle_{\omega_k}=
-\sum_I \frac{f^{yx}_I}{\omega_I^2-\omega_k^2},
\end{array}
\end{eqnarray}
where the oscillator strengths are determined as follows 
\begin{eqnarray}\label{osc_strengths}\begin{array} {l}
\displaystyle
{f^{yx}_I}=2\omega_I\big(X^\dagger(\omega_I)V^{[1]\mu}_{y}\big)^\dagger
X^\dagger(\omega_I)V^{[1]\mu}_{x}.
\end{array}
\end{eqnarray}
The last three equations (Eqs.~\ref{linear_rsp_function}--\ref{osc_strengths}) comprise
 the ingredients for calculation of excitation energies and intensities within the TD-MC-srDFT scheme. 

\subsection{SOPPA-srDFT model}\label{subsec:soppa-srdft}
As an alternative to TD-MC-srDFT for systems which are not strongly
multi-configurational, the SOPPA-srDFT scheme will now be introduced.
It consists of an application of the SOPPA
approach~\cite{Oddershede84_soppa,packer96_soppa} to the auxiliary long-range
interacting system. For that purpose we will replace in the TD-MC-srDFT
linear response Eq.~(\ref{lr_srdft_eq_minusomega}) the unperturbed MC-srDFT wave function
$\vert 0^\mu\rangle $ by a M{\o}ller-Plesset (MP)
perturbation expansion through second order in the long-range
fluctuation potential~\cite{fromager2008,mp2srdftdmat}   
\begin{eqnarray}\label{MP2-srdft_wfexp}\begin{array} {l}
\displaystyle
\vert{0}^{\mu}\rangle\rightarrow \vert{\rm
HF}^{\mu}\rangle+\vert{0}^{(1)\rm
lr,\mu}\rangle+\vert{0}^{(2)\mu}\rangle+\ldots,
\end{array}
\end{eqnarray}
where $\vert{\rm HF}^{\mu}\rangle$ denotes the HF-srDFT determinant. The first-order
contribution is the analog of the standard MP1 
wave function correction 
based on the long-range Hamiltonian
$\hat{H}^{\mu}[\rho^\mu_{\mbox{\tiny HF}}]$
that is calculated
for the HF-srDFT density $\rho^\mu_{\mbox{\tiny HF}}$
, while the second-order term includes self-consistency
effects~\cite{fromager2008}. Based on the analysis and
numerical results of Fromager and Jensen~\cite{mp2srdftdmat}, where it was shown that
these effects can be safely neglected through second order,
self-consistency will be not be included in the presented SOPPA-srDFT
results. According to the Brillouin theorem the density remains unchanged through
first order which explains why self-consistency only appears through
second order in the wave function. The density can therefore be expanded
as 
\begin{eqnarray}\label{MP2-srdft_denexp}\begin{array} {l}
\displaystyle
\rho_0^{\mu}(\mathbf{r})\rightarrow
 \rho^{\mu}_{\rm HF}(\mathbf{r})+\delta\rho^{(2)\mu}(\mathbf{r})+\ldots
\end{array}
\end{eqnarray}
The SOPPA-srDFT equations are then obtained when expanding the linear
response Eq.~(\ref{lr_srdft_eq_minusomega}) through second order in the long-range fluctuation
potential. The $\vert D_i \rangle\langle \text{HF}^{\mu}\vert$ operator corresponds to what in the original SOPPA literature is denoted the 
two-particle-two-hole operator. Since one and two particle-hole manifolds are sufficient
to define the SOPPA response~\cite{packer96_soppa}, 
the orbital and configuration
rotation operators can be written as 
\begin{eqnarray}\label{orb-conf-rot-soppa}
\hat{q}^\dagger_i &\rightarrow &\hat{E}_{ai}
\nonumber
\\
\hat{R}^\dagger_i &\rightarrow &\vert D_i\rangle\langle {\rm
HF}^\mu\vert,
\end{eqnarray}
where $i$ and $a$ are
occupied and unoccupied HF-srDFT orbitals, respectively, while $\vert D_i\rangle$
denote singlet and triplet doubly-excited states.
Since the metric and the gradient property vector 
in Eq.~(\ref{lr_srdft_eq_minusomega}) depend on the wave function
through expectation values only (see Eqs.~(\ref{metricmu}), (\ref{SDmu}) and
(\ref{gradientdensvector})), their
expressions in SOPPA-srDFT are obtained from standard SOPPA when
replacing the regular Hamiltonian by
$\hat{H}^{\mu}[\rho^\mu_{\mbox{\tiny HF}}]$, as self-consistency effects
on the wave function are neglected through second order. 

The derivation of the Hessian 
requires more discussion as it also depends on the density through the srHxc
potential and kernel. 
Note that, in
order to obtain the correct linear response function through second
order, the Hessian matrix
elements should be computed through second order in the orbital-orbital
blocks, first order in the orbital-configuration blocks and zeroth
order in the configuration-configuration blocks~\cite{packer96_soppa}. 
According to Eq.~(\ref{MP2-srdft_denexp}), the
long-range interacting Hamiltonian in Eq.~(\ref{ABmu}) is expanded through second order as follows
\begin{eqnarray}\label{srHxc_pot_pt2}
\displaystyle
\hat{H}_0^{\mu}&\rightarrow&\hat{H}^{\mu}[\rho^\mu_{\mbox{\tiny HF}}]
\nonumber
\\
&&+ 
\int\int {\rm d}\mathbf{r}{\rm d}\mathbf{r'}\; K_{\rm Hxc}^{\rm
sr,\mu}[\rho^\mu_{\mbox{\tiny HF}}](\mathbf{r},\mathbf{r'})\delta
\rho^{(2)\mu}(\mathbf{r})\,\hat{\rho}(\mathbf{r'})
\nonumber
\\
&&+\ldots
\end{eqnarray}
The second-order correction in
Eq.~(\ref{srHxc_pot_pt2}) needs to be considered in the
orbital-orbital blocks of the Hessian only, 
leading to the following
contribution for the upper left block  
\begin{eqnarray}\label{orb-orb_hessian_PT2}
&&
\int\int {\rm d}\mathbf{r}{\rm d}\mathbf{r'}\; K_{\rm Hxc}^{\rm
sr,\mu}[\rho^\mu_{\mbox{\tiny HF}}](\mathbf{r},\mathbf{r'})\delta
\rho^{(2)\mu}(\mathbf{r})
\nonumber
\\
&&\times\langle
{\rm HF}^{\mu}\vert[\hat{E}_{ia},[\hat{\rho}(\mathbf{r'}),\hat{E}_{bj}]]\vert
{\rm HF}^{\mu}\rangle
\nonumber
\\
&=&
2\int\int {\rm d}\mathbf{r}{\rm d}\mathbf{r'}\; K_{\rm Hxc}^{\rm
sr,\mu}[\rho^\mu_{\mbox{\tiny HF}}](\mathbf{r},\mathbf{r'})\delta
\rho^{(2)\mu}(\mathbf{r})
\nonumber
\\
&&\hspace{0.6cm}\times
\Big(\delta_{ij}\Omega_{ab}(\mathbf{r'})-\delta_{ab}\Omega_{ij}(\mathbf{r'})\Big),
\end{eqnarray}
where $\Omega_{pq}(\mathbf{r})=\phi_p(\mathbf{r})\phi_q(\mathbf{r})$
denotes 
the product of HF-srDFT orbitals.
Let us now consider the srHxc kernel contribution to the Hessian in
Eq.~(\ref{srkernelHessian}) that is
determined from the
following perturbation expansion through
second order 
\begin{eqnarray}\label{srHxc_kern_pt2}
K_{\rm Hxc}^{\rm sr,\mu}[\rho_0^\mu](\mathbf{r},\mathbf{r'})
&\rightarrow& 
K_{\rm Hxc}^{\rm sr,\mu}[\rho^\mu_{\mbox{\tiny HF}}](\mathbf{r},\mathbf{r'})
\nonumber
\\
&&+\int {\rm d}\mathbf{r''}\; 
\frac{\delta K_{\rm Hxc}^{\rm sr,\mu}}
{\delta \rho(\mathbf{r''})}[\rho^\mu_{\mbox{\tiny HF}}]
(\mathbf{r},\mathbf{r'})
\delta
\rho^{(2)\mu}(\mathbf{r''})
\nonumber
\\
&&+\ldots
\end{eqnarray}
The second-order term in Eq.~(\ref{srHxc_kern_pt2}) should be considered
in the orbital-orbital blocks only, 
leading to the following
contribution in the upper left block  
\begin{eqnarray}\label{srHxc_kern_pt2_orb-orb}
&&
\int \int\int{\rm d}\mathbf{r}{\rm d}\mathbf{r'} {\rm d}\mathbf{r''}\; 
\frac{\delta K_{\rm Hxc}^{\rm sr,\mu}}
{\delta \rho(\mathbf{r''})}[\rho^\mu_{\mbox{\tiny HF}}]
(\mathbf{r},\mathbf{r'})
\delta
\rho^{(2)\mu}(\mathbf{r''})
\nonumber
\\
&&\times 
\langle{\rm HF}^{\mu}\vert[\hat{E}_{ia},\hat{\rho}(\mathbf{r})]\vert{\rm
HF}^{\mu}\rangle
\langle{\rm HF}^{\mu}\vert[\hat{\rho}(\mathbf{r'}),\hat{E}_{bj}]\vert{\rm
HF}^{\mu}\rangle
\nonumber
\\
\nonumber
\\
&&=
4\int\int\int{\rm d}\mathbf{r}{\rm d}\mathbf{r'} {\rm d}\mathbf{r''}\; 
\frac{\delta K_{\rm Hxc}^{\rm sr,\mu}}
{\delta \rho(\mathbf{r''})}[\rho^\mu_{\mbox{\tiny HF}}]
(\mathbf{r},\mathbf{r'})
\delta
\rho^{(2)\mu}(\mathbf{r''})
\nonumber
\\
&&\hspace{3.2cm}\times\Omega_{ai}(\mathbf{r})\Omega_{bj}(\mathbf{r'}).
\end{eqnarray}
The remaining contributions to the Hessian that have to be considered arise from the srHxc kernel
calculated with the HF-srDFT density 
\begin{eqnarray}\label{srkernelHessian_HF-srDFT}\begin{array} {l}
\displaystyle
\int\int {\rm d}\mathbf{r}{\rm d}\mathbf{r'}\; K_{\rm Hxc}^{\rm
sr,\mu}[\rho_{\mbox{\tiny
HF}}^\mu](\mathbf{r},\mathbf{r'})\rho^{[1]\mu}(\mathbf{r})\rho^{[1]\mu\dagger}(\mathbf{r'}),
\end{array}
\end{eqnarray}
where the perturbation expansion of the orbital components in the gradient density vector 
\begin{eqnarray}\label{orbcompt_1RDM}
\langle0^{\mu}\vert[\hat{E}_{ia},\hat{\rho}(\mathbf{r})]\vert0^{\mu}\rangle
&=&\sum_{p,q}\Omega_{pq}(\mathbf{r})
\langle0^{\mu}\vert[\hat{E}_{ia},\hat{E}_{pq}]\vert0^{\mu}\rangle
\nonumber
\\
&=&\sum_{p,q}\Omega_{pq}(\mathbf{r})
\nonumber
\\
&&\hspace{0.7cm}\times\Big(\delta_{ap}D^{\mu}_{iq}-\delta_{iq}D^{\mu}_{pa}\Big),
\end{eqnarray}
is deduced from the {one-electron reduced density matrix} (1RDM) expansion
\begin{eqnarray}\label{1RDMpt2}
D^{\mu}_{pq}&=&
\langle0^{\mu} \vert\hat{E}_{pq}\vert0^{\mu}\rangle
\nonumber\\
&\rightarrow&
\sum_i 2\delta_{ip}\delta_{iq}
+D_{pq}^{(2)\mu}+\ldots
\end{eqnarray}
Note that the first-order contribution to the 1RDM is zero because of
the Brillouin theorem~\cite{fromager2008}. We thus obtain through second order 
\begin{eqnarray}\label{orbcompt_1RDM_PT2}
&&\langle0^{\mu}\vert[\hat{E}_{ia},\hat{\rho}(\mathbf{r})]\vert0^{\mu}\rangle
\rightarrow2\Omega_{ai}(\mathbf{r})
\nonumber
\\
\nonumber
\\
&&+\sum_{p}\Big(\Omega_{pa}(\mathbf{r})
D^{(2)\mu}_{ip}-\Omega_{pi}(\mathbf{r})D^{(2)\mu}_{pa}\Big)
+\ldots
\end{eqnarray}
It was shown numerically by Fromager and Jensen~\cite{mp2srdftdmat}
that, for the usual $\mu=0.4$ value, the long-range MP2 contribution to
the 1RDM is relatively small as long
as it is computed for systems that are not strongly
multi-configurational. As a result, the second-order contributions in
Eqs.~(\ref{orb-orb_hessian_PT2}), (\ref{srHxc_kern_pt2_orb-orb}) and
(\ref{orbcompt_1RDM_PT2}) have
been neglected in our implementation 
Let us finally focus on the 
configuration part of the gradient density vector  
that must be expanded through first order in order to compute the
srHxc kernel orbital-configuration blocks: 
\begin{eqnarray}\label{confcompt_PT1}
\langle0^{\mu}\vert[\hat{R}_{i},\hat{\rho}(\mathbf{r})]\vert0^{\mu}\rangle
&\rightarrow&\langle{\rm
HF}^{\mu}\vert[\hat{R}_{i},\hat{\rho}(\mathbf{r})]\vert
{0}^{(1)\rm lr,\mu}\rangle
+\ldots
\nonumber
\\
&=&\langle D_i\vert \hat{\rho}(\mathbf{r})\vert {0}^{(1)\rm
lr,\mu}\rangle
\nonumber
\\
&&-\rho^\mu_{\rm HF}(\mathbf{r})\langle
D_i\vert{0}^{(1)\rm lr,\mu}\rangle+\ldots
\end{eqnarray}
Rewriting the long-range MP1 wave function in the basis of the doubly-excited configurations  
\begin{eqnarray}\label{MP1wf}
\vert
{0}^{(1)\rm lr,\mu}\rangle=\sum_jC_j^{(1)\rm lr,\mu}\vert D_j\rangle,
\end{eqnarray}
we obtain
\begin{eqnarray}\label{confcompt_PT1_2}
\langle0^{\mu}\vert[\hat{R}_{i},\hat{\rho}(\mathbf{r})]\vert0^{\mu}\rangle
&\rightarrow&
C_i^{(1)\rm lr,\mu}\Big(
\langle D_i\vert \hat{\rho}(\mathbf{r})\vert D_i\rangle -\rho^\mu_{\rm
HF}(\mathbf{r})\Big)
\nonumber
\\
&&+\sum_{j\neq i}C_j^{(1)\rm lr,\mu}\langle D_i\vert
\hat{\rho}(\mathbf{r})\vert D_j\rangle.
\end{eqnarray}
This term may contribute significantly to the Hessian when considering double excitations
with an important modification of the density. 
For simplicity it has been neglected in this work.
In summary, the SOPPA-srDFT equation that has been implemented has the same structure as the linear response
TD-MC-srDFT equation. 
The long-range interacting Hessian $E_0^{[2]\mu}$ has been replaced by the SOPPA analog based on
$\hat{H}^{\mu}[\rho^\mu_{\mbox{\tiny HF}}]$ while the srHxc kernel
contribution has been calculated for the HF-srDFT density with the gradient
density vector simplified as follows, according to
Eq.~(\ref{orbcompt_1RDM_PT2}),
\begin{eqnarray}\label{gradientdensvector_soppa-srdft}\begin{array} {l}
\rho^{[1]\mu}(\mathbf{r})\rightarrow \begin{bmatrix}
2\Omega_{ai}(\mathbf{r})\\
0\\
-2\Omega_{ai}(\mathbf{r})\\
0
\end{bmatrix}
.
\end{array}
\end{eqnarray}
\section{\label{Compdetails} Computational Details}       
 
\begin{figure*}[htb!]
 \centering
 \includegraphics[height=4.5cm]{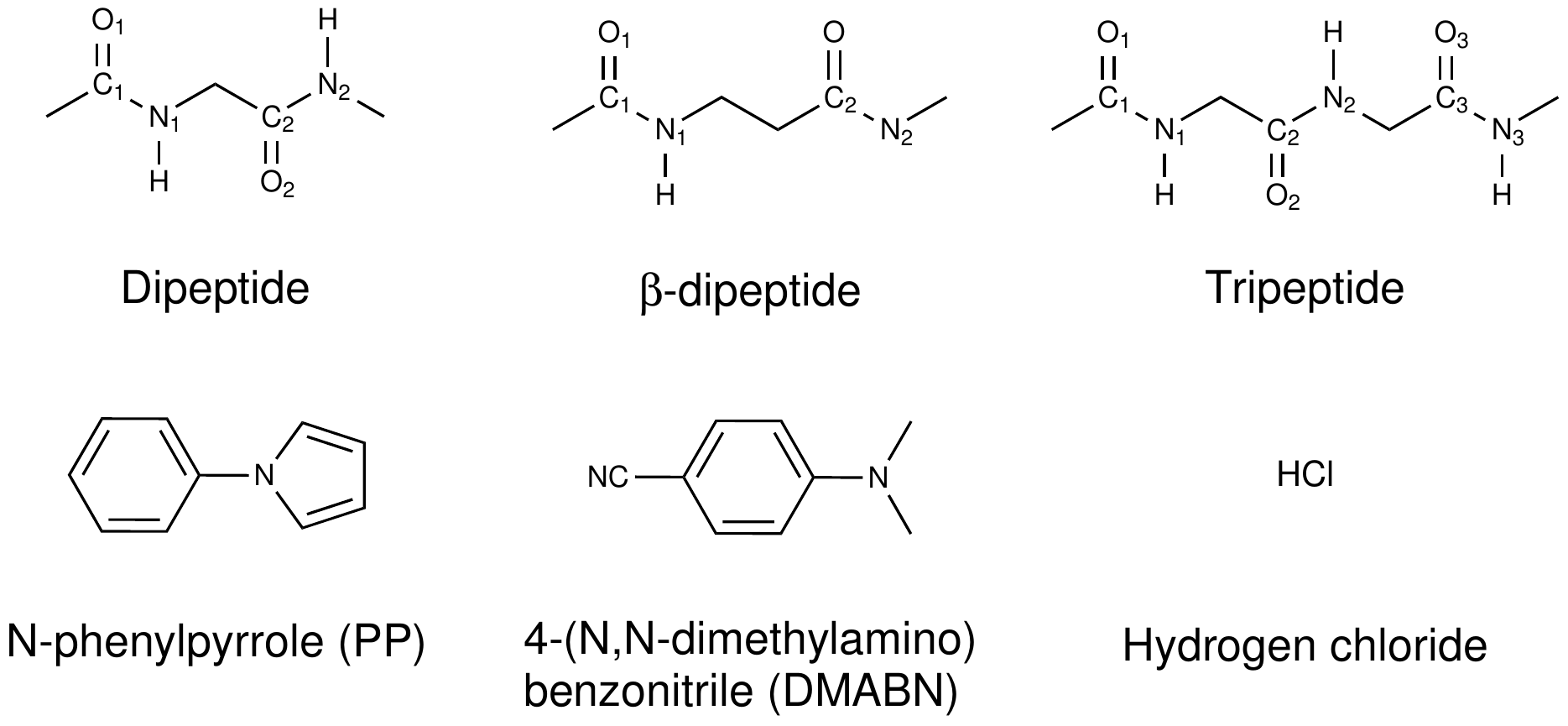}
 \caption{Molecules used for statistical analysis.\label{molecules}}
\end{figure*} 
The set of molecules used to benchmark the TD-MC-srDFT method is shown in Figure \ref{molecules}. 
It comprises hydrogen chloride, three model peptides (a simple dipeptide, a $\beta$-dipeptide and  
a tripeptide) and two aromatic systems, $N$-phenyl pyrrole (PP) and 4-($N,N$-dimethylamino) benzonitrile (DMABN). 
Excitations of local and charge transfer character for these systems have previously been investigated, 
as part of a larger 
test set introduced by Peach \textit{et al.}\cite{peach2008} for benchmarking
the three-parameter Becke-Lee-Yang-Parr functional (B3LYP) and its long-range corrected version. The long-range correction has the form of
the Coulomb-attenuated method and is hence denoted CAM-B3LYP\cite{yanai2004}.   
Many of the CT excitations in the chosen systems have been problematic for TD-DFT\cite{tozer1999,xu2006,proppe2000}, 
but can be improved with long-range corrected functionals. 
In order to assure a fair comparison of our data with the values used to
benchmark CAM-B3LYP, we took the geometries from the study by Peach \textit{et al.}\cite{peach2008} 
As reference for the calculated excitation energies, we use for the dipeptide, $\beta$-dipeptide and tripeptide,
the CASPT2 values from Serrano-Andr{\'e} and F{\"u}lscher\cite{serrano1998}. 
Excitation energies for $N$-phenyl pyrrole (PP) and HCl were obtained at
the linear response CC level by Peach \textit{et al.}\cite{peach2008}
Finally, reference excitation data for DMABN were taken from accurate gas phase measurements\cite{bulliard1999}, 
thus obtaining a one-to-one correspondence between the reference values used by Peach \textit{et al.} 
and the ones adopted here. DMABN has been subject to several theoretical studies\cite{rappoport2004,serrano1995} 
and CASPT2(12,12) reference values could alternatively have been used.
In our calculations we use CAS(4,4) spaces for the dipeptide and $\beta$-dipeptide, while  
for the tripeptide, a slightly larger CAS(6,6) active space was applied. The two organic molecules (PP and DMABN) are both assigned CAS(8,8) active spaces.
All calculations for the molecules in Figure \ref{molecules} are carried
out with a Dunning cc-pVTZ basis set\cite{bs890115d}. The srDFT
calculations were performed with the 
spin-independent short-range exchange-correlation functional of Goll
{\it et al.}~\cite{goll2005} which is based on the
Perdew-Burke-Ernzerhof (PBE) functional. It will therefore be referred
to as srPBE. The $\mu$ parameter was set to $\mu=0.4$. 
This value relates to a prescription given in
Refs.~\cite{fromager2007,fromager2009} where $\mu=0.4$ was found optimal, based on an analysis of correlation
effects in the MC-srPBE ground state. 

\begin{figure}[htb!]
 \centering
 \includegraphics[height=1.5cm]{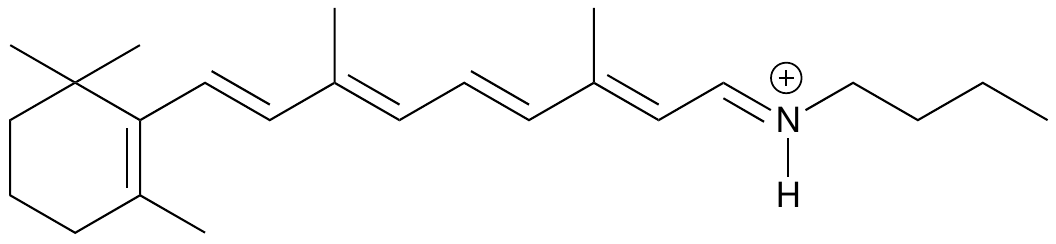}
 \caption{Retinal chromophore (including a small part of the lysine residue which attach retinal to the channel-rhodopsin protein).\label{retinal_chromophore}}
\end{figure}

The SOPPA and SOPPA-srDFT calculations were not done for the full set, 
but only for the smallest model peptide (dipeptide), PP and DMABN molecules. 
The retinal chromophore is in the all-\textit{trans} Schiff-base form (see
Figure \ref{retinal_chromophore}). We used a structure from a very recent study\cite{sneskov2013}, optimized within the protein environment (using B3LYP/6-31+G*). 
For this system a CAS(6,6) space was chosen based on MP2-srPBE natural orbital occupation numbers~\cite{mp2srdftdmat}. 
TD-srPBE calculations on retinal were performed with a 6-31G* basis set, and we have accordingly not included the results from the retinal calculations 
in any of the statistical analysis presented in Section \ref{results}.  

All calculations were carried out using a development version of the DALTON program\cite{DALTON2011}.

\section{\label{results}Results and Discussion}



\subsection{\label{res2}Classification of Excitations}


To identify whether a given excitation is of local (``L'') or of charge transfer (``CT'') character, 
response vectors from the calculation of excitation energies for each of the molecular systems in Figure \ref{molecules} have been through a careful analysis. 
This includes analysis of both transitions between orbitals and
configurations along with visual inspection of the orbitals involved.
Results are given in Table 
\ref{SOPPA} for the dipeptide, PP and DMABN molecules. Table \ref{TD-MC-srDFTtable} shows TD-MC-srPBE results for the remaining
molecules ($\beta$-dipeptide, tripeptide and HCl). We here mainly discuss the excitations which are qualitatively different from previous benchmark results and accordingly  
the dipeptide, DMABN, HCl and PP molecules (which all give the same
qualitative excitation profile as previous calculations) will not be
discussed in detail: Focus will be put on 
the $\beta$-dipeptide and tripeptide, but all excitation energies and
assignments are included in the supporting information\cite{supporting}.   
\begin{table*}[htb!]
\centering
\caption{Vertical excitation energies (in eV). 
"sr" is shorthand for "srPBE" 
\label{SOPPA}}
 \begin{tabular}{lllcccccccc}
    \hline \hline \\[-1.5ex]
 Molecule  & Assign.                                & Type    & TD-HF      & TD-HF-sr &  SOPPA & SOPPA-sr & TD-MC-sr & TD-B3LYP & TD-CAM-B3LYP & Ref \\[0.5ex]
    \hline \\[-1.5ex] 
Dipeptide  & $n_{1}\rightarrow\pi^{*}_{1}$          & L      & 6.55     &  6.41      & 5.01  & 5.53  &  5.61    & 5.55  & 5.68                & 5.62$^{a}$\tnote{a} \\[0.5ex]
Dipeptide  & $n_{2}\rightarrow\pi^{*}_{2}$          & L      & 6.78     &  6.64      & 5.22  & 5.75  &  5.83    & 5.77  & 5.92                & 5.79$^{a}$\tnote{a} \\[0.5ex]
Dipeptide  & $\pi_{1}\rightarrow\pi^{*}_{2}$        & CT$_1$ & 8.44     &  7.47      & 6.58  & 6.95  &  7.59    & 6.15  & 7.00               & 7.18$^{a}$\tnote{a} \\[0.5ex]
Dipeptide  & $n_{1}\rightarrow\pi^{*}_{\text{N2}}$  & CT$_2$ & 8.98     &  7.60      & 6.85  & 7.10  &  8.10    & 6.31  & 7.84               & 8.07$^{a}$\tnote{a} \\[0.5ex] 
PP         & $\pi_{1}\rightarrow\pi^{*}_{1}$        & L      & 5.83     &  5.20      & 4.26  & 4.97  &  5.40    & 4.76  & 5.06               & 4.85$^{b}$\tnote{b} \\[0.5ex]
PP         & $\pi_{2}\rightarrow\pi^{*}_{2}$        & L      & 5.41     &  5.31      & 4.55  & 5.04  &  5.48    & 4.96  & 5.12               & 5.13$^{b}$\tnote{b} \\[0.5ex]
PP         & $\pi_{1}\rightarrow\pi^{*}_{2}$        & CT     & 5.57     &  5.68      & 4.99  & 5.25  &  5.70    & 4.58  & 5.27               & 5.47$^{b}$\tnote{b} \\[0.5ex]
PP         & $\pi_{2}\rightarrow\pi^{*}_{1}$        & CT     & 7.40     &  6.89      & 5.57  & 5.84  &  6.65    & 4.64  & 5.92               & 5.94$^{b}$\tnote{b} \\[0.5ex]
DMABN      & $\pi_{1}\rightarrow\pi^{*}_{\text{CN}}$ & L     & 5.41     &  4.88      & 3.87  & 4.56  &  5.09    & 4.44  & 4.72      & 4.25$^{c}$\tnote{c} \\[0.5ex]
DMABN      & $\pi_{2}\rightarrow\pi^{*}_{\text{CN}}$ & CT    & 5.22     &  5.06      & 4.14  & 4.73  &  5.12    & 4.64  & 4.91      & 4.56$^{c}$\tnote{c} \\[0.5ex]
    \hline \hline
 \end{tabular} 
\centering
 \begin{tablenotes}\footnotesize 
\item[a] \hspace{15.0mm}$^{a}$CASPT2 results from Serrano-Andr{\'e} and F{\"u}lscher\cite{serrano1998}
 \item[b] \hspace{15.0mm}$^{b}$CC2 results from Peach \textit{et al.}\ \cite{peach2008}
 \item[c] \hspace{15.0mm}$^{c}$Gas phase experiment from Bulliard \textit{et al.}\ \cite{bulliard1999}
 \end{tablenotes}
\end{table*}
The model peptides  display 
local excitations of $n\rightarrow\pi$ character within the carbonyl groups 
and of $\pi\rightarrow\pi^{*}$ character between the carbonyl and peptide bonds (denoted ``W'' and ``NV'' in ref.\citenum{serrano1998}). 
Seeing that the $\pi\rightarrow\pi^{*}$ ``NV'' type of excitations have been left out from the study of Peach \textit{et al.}\cite{peach2008} 
we will only briefly discuss them here. These excitations have also been removed from Table \ref{SOPPA} (see ref.~\citenum{supporting}) 
and they will not be included in the statistical analysis in Section \ref{srMCSCFpeform}.
 The peptide models further display two different kinds of charge transfer excitations: 
either these involve the peptide $\pi\rightarrow\pi^{*}$ systems (``CT$_1$'')  or 
the carbonyl $n\rightarrow\pi^{*}$ lone pairs (``CT$_2$''). 
\begin{table*}[htb!]
\centering
\caption[S1]{Vertical excitation energies (in eV). 
"sr" is shorthand for "srPBE" 
\label{TD-MC-srDFTtable}}
 \begin{tabular}{llllccccccc}
    \hline \hline \\[-1.5ex]
 Molecule       & Assign.                           &  Assignment  & TD-MC-sr  & TD-B3LYP & TD-CAM-B3LYP & Ref \\[0.5ex]
    \hline \\[-1.5ex]
$\beta$-Dipeptide & $n_{2}\rightarrow \pi^{*}_{2}$       & L     &  5.60     & 5.56  & 5.67      & 5.40$^{a}$\tnote{a}   \\[0.5ex]
$\beta$-Dipeptide & $n_{1}\rightarrow \pi^{*}_{1}$       & L      &  5.74     & 5.66  & 5.76      & 5.10$^{a}$\tnote{a}   \\[0.5ex]
$\beta$-Dipeptide & $\pi_{1}\rightarrow \pi^{*}_{\text{N}2}$ & CT$_1$  &  7.41     &  7.2  &   8.01    & 7.99$^{a}$\tnote{a}   \\[0.5ex]
$\beta$-Dipeptide & $n_{1}\rightarrow \pi^{*}_{2}$           & CT$_2$      &  8.21     & 7.26  &   8.38    & 9.13$^{a}$\tnote{a}   \\[0.5ex]
Tripeptide        &  $n_{1}\rightarrow\pi^{*}_{1}$       & L     &   5.66    &  5.57  & 5.72     & 5.74$^{a}$\tnote{a}      \\[0.5ex]
Tripeptide        & $n_{3}\rightarrow\pi^{*}_{3}$        & L     &   5.87    &  5.74  & 5.93     & 5.61$^{a}$\tnote{a}      \\[0.5ex]
Tripeptide        & $n_{2}\rightarrow\pi^{*}_{2}$        & L     &   5.92    &  5.88  & 6.00     & 5.92$^{a}$\tnote{a}      \\[0.5ex]
Tripeptide        & $\pi_{1}\rightarrow\pi^{*}_{2}$      &  CT$_1$&  8.12  &  6.27        &  6.98  & 7.01$^{a}$\tnote{a}  \\[0.5ex]
Tripeptide        & $\pi_{2}\rightarrow \pi^{*}_{3}$     & CT$_1$ &  8.31  &  6.60        &  7.69  & 7.39$^{a}$\tnote{a}  \\[0.5ex]
Tripeptide        & $\pi_{1}\rightarrow \pi^{*}_{\text{N}3}$ & CT$_1$& 8.43&  6.06        &  8.51  & 8.74$^{a}$\tnote{a}  \\[0.5ex]
Tripeptide        & $n_{1}\rightarrow\pi^{*}_{\text{N}2}$ & CT$_2$ & 8.52  &  6.33        &  7.78  & 8.12    \\[0.5ex]
Tripeptide        & $n_{2}\rightarrow\pi^{*}_{\text{N}3}$ & CT$_2$ & 8.84  &  6.83        &  8.25  & 8.33    \\[0.5ex]
Tripeptide        & $n_{1}\rightarrow\pi^{*}_{3}$          & CT$_2$ & 9.04  &  6.12        &  8.67  & 9.30    \\[0.5ex]
HCl                & $\, ^{1}\Pi$                         & CT & 8.03     & 7.65  & 7.79      & 8.23$^{b}$\tnote{b}    \\[0.5ex]       
\hline \hline
 \end{tabular} 
\centering
 \begin{tablenotes}\footnotesize 
\item[a] \hspace{15.0mm}$^{a}$CASPT2 results from Serrano-Andr{\'e} and F{\"u}lscher, ref.~\citenum{serrano1998}
 \item[b] \hspace{15.0mm}$^{b}$CC2 results from Peach \textit{et al.}\ \cite{peach2008}
 \end{tablenotes}
\end{table*}
In the dipeptide the local carbonyl $n_{1}\rightarrow\pi^{*}$ and $n_{2}\rightarrow\pi^{*}$ excitations occur in the same order for TD-MC-srPBE and CASPT2. However, 
this changes for the $\beta$-dipeptide where the two local transitions
occur in reversed order at the TD-MC-srPBE level, compared to the CASPT2
results.\cite{supporting} 
The reversed ordering of these two excitations in the $\beta$-dipeptide corresponds to what is obtained by B3LYP and CAM-B3LYP functionals. 
We note that the inversion of excitations in the $\beta$-peptide also
occurs for $\pi\rightarrow\pi^{*}$ type of excitations 
(``NV$_{1}(1)$'' and ``NV$_{1}(2)$'' in ref.~\citenum{serrano1998}). 
We thus obtain NV$_{1}(2)$ as the lowest of the two excitations. 
\begin{figure}[htb!]
   \includegraphics[height=2.4cm]{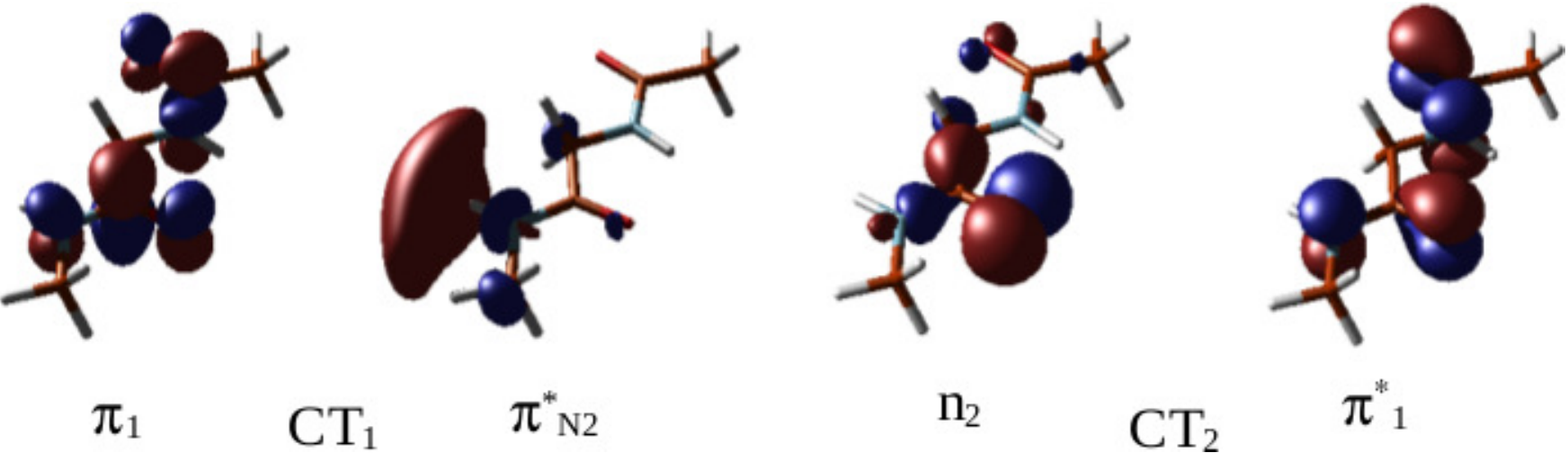}
 \caption{Orbitals involved in the two  charge transfer excitations in the dipeptide.\label{chargetransferorbs_2}}
\end{figure}
In the tripeptide, the lowest TD-MC-srPBE excitations are the local intra carbonyl excitations, 
which is in agreement with the reference CASPT2 results, but the order
of the two first excitations are again reversed. This inversion is also
observed at both 
TD-B3LYP and TD-CAM-B3LYP levels. 
The three NV$_1$ excitations seem to come in the same order as in CASPT2, although this cannot be unequivocally verified since the first two are nearly degenerate.
A word of caution is also necessary for the classification of the charge transfer excitations. 
We find that the accepting orbitals occasionally are mainly located at the peptide \ce{N-H} bond as shown in Figure \ref{chargetransferorbs_2} (using the dipeptide as example).
 These kinds of accepting orbitals are denoted ``$\pi_{\text{N}}$''
 orbitals in Table \ref{SOPPA} and Table \ref{TD-MC-srDFTtable}.
It should finally be mentioned that we seem to experience more mixing of
states in our TD-MC-srPBE calculations than in the reference
CASPT2 calculations\cite{serrano1998}. One reason for this might be that 
the CASPT2 benchmarks were performed with an ANO type basis of double zeta quality, whereas we have used the more extensive cc-pVTZ. 
The use of different sized basis sets might also be the reason for the inversion of states described above. 
\begin{figure}[htb!]
 \centering
\includegraphics[height=6cm]{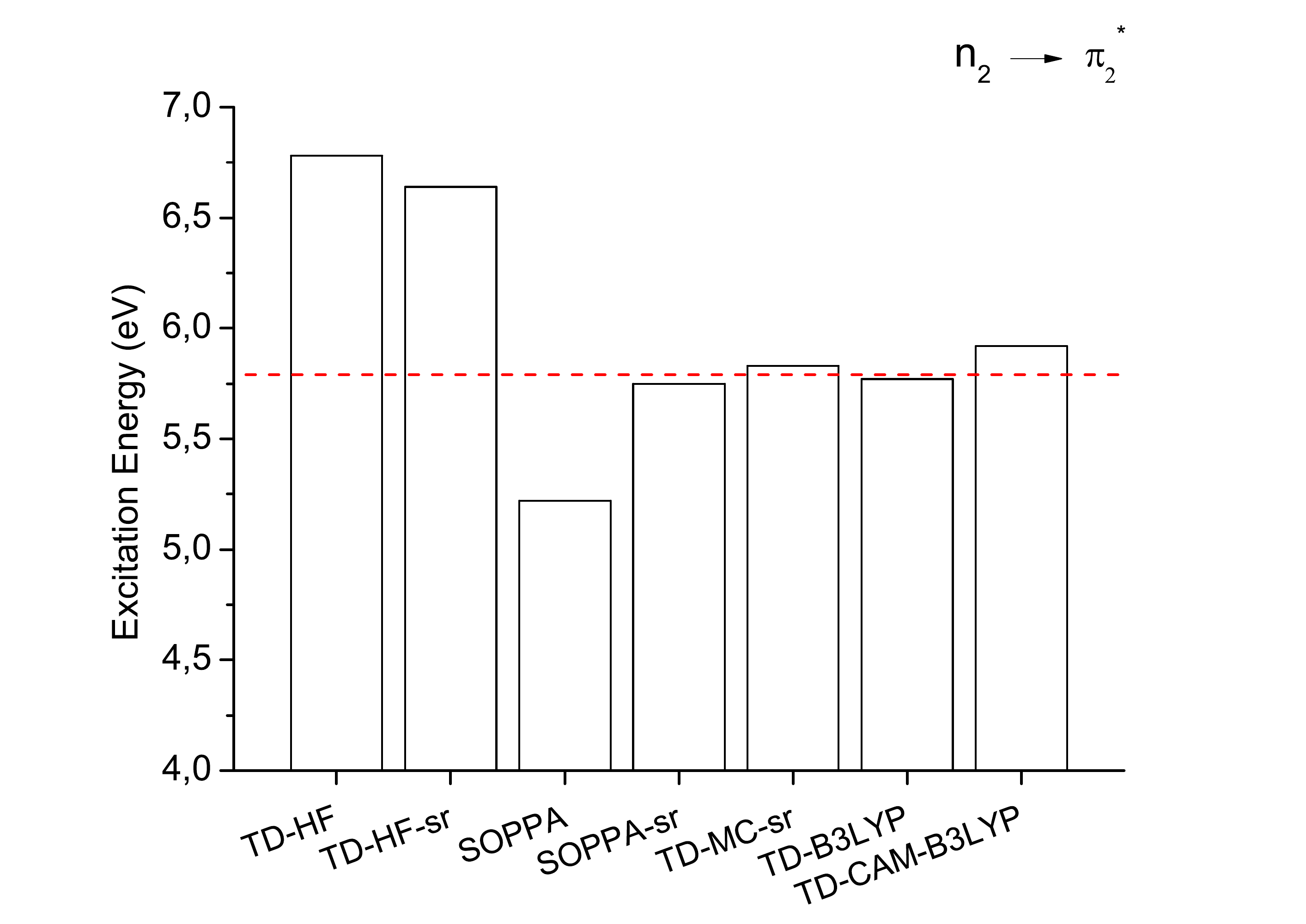}
 \caption{Local excitation in the model dipeptide. The red dotted line is the CASPT2 results from ref.~\citenum{serrano1998}. ``sr'' is short-hand for ``srPBE''.\label{dipepW2}}
\end{figure}
\begin{figure}[htb!]
 \centering
 \includegraphics[height=6cm]{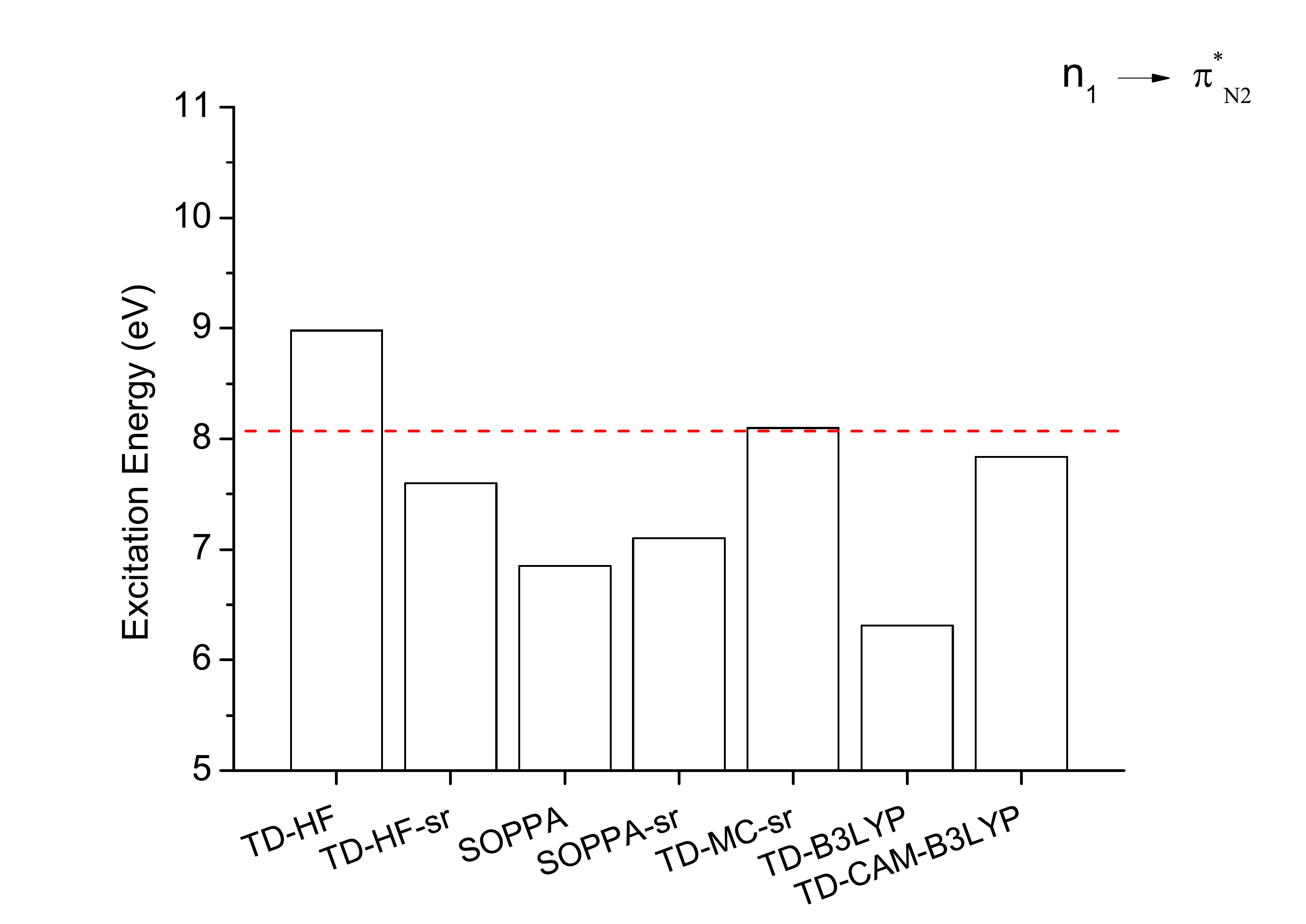}
 \caption{Charge transfer excitation in the model dipeptide. The red dotted line is the CASPT2 results from ref.~\citenum{serrano1998}.  ``sr'' is short-hand for ``srPBE''.\label{dipepCT2}}
\end{figure}

\subsection{Comparing SOPPA-srPBE with TD-MC-srPBE results}

The subset of molecules considered in this section is given in Table \ref{SOPPA}, where also the results are compiled. 
For the dipeptide, the performance of the various methods is depicted in
Figures \ref{dipepW2}\ and \ref{dipepCT2}\ for the $n_{2}\rightarrow\pi^{*}_{2}$ (``W$_{2}$'') and the charge transfer (``CT$_2$'') transitions, respectively.  
As expected the local transitions are overestimated at the TD-HF level and 
we expect a similar situation within regular TD-MCSCF (a good estimate for this overestimation is provided by considering
\textit{e.g.}\ the results from state-averaged CASSCF from ref.~\citenum{rappoport2004} which is about 1.4 eV
too high for the DMABN molecule). 
TD-HF-srPBE  leads to a change in the right direction, but it is not
sufficient to obtain agreement with the reference CASPT2 values.   
SOPPA significantly underestimates the local carbonyl excitations while 
SOPPA-srPBE is very close to the CASPT2 values for the local
$n_{1}\rightarrow\pi^{*}_{1}$ (``W$_1$'') and $n_{2}\rightarrow\pi^{*}_{2}$ (``W$_2$'') excitations. 
The TD-MC-srPBE model also remedies the tendency 
to overestimate excitation energies from the MCSCF type of wave functions and the two local excitations are obtained very accurately.
Both TD-B3LYP and TD-CAM-B3LYP are, as expected, also of high accuracy for these two excitations. 
Moving to the charge transfer excitations, both Table \ref{SOPPA} and Figure \ref{dipepCT2} show 
that these are severely underestimated by the B3LYP functional in the dipeptide.
The CAM-B3LYP functional provides slightly better results, 
which is not surprising as its parameters have been optimized for
reproducing such excitations well. 
 TD-HF and SOPPA behave similarly as for the local excitations and thus 
overestimate and underestimate, respectively, the charge transfer excitations. 
Note that, 
for the charge transfer ``CT$_2$'', TD-HF-srPBE is closer to CASPT2 than
SOPPA-srPBE, even though the latter performs better than SOPPA. 
The TD-MC-srPBE method is also for charge transfer excitations very accurate and for the dipeptide it even outperforms
CAM-B3LYP. 

We can from the discussion for the dipeptide also comment on some general trends in Table \ref{SOPPA}. 
As documented many times before, 
TD-HF overestimates both charge transfer and local excitations and the TD-HF-srPBE method generally brings the result closer to the reference data. 
However, the correspondence is still not satisfactory for the method to
be of use for quantitative treatments, as it neglects long-range
correlation effects. 
The regular SOPPA model generally underestimates both local and charge transfer excitations, while 
the SOPPA-srPBE method is a significant improvement for both types of excitations in all molecules considered. 
The TD-MC-srPBE method is often an improvement compared to TD-HF and also to TD-HF-srPBE. 
For charge transfers, TD-MC-srPBE is in general also an improvement to B3LYP and occasionally even to CAM-B3LYP. 
One notable exception is the DMABN molecule, where B3LYP previously has been noted to perform well, also for charge-transfer excitations\cite{jamorski2002}. 
In the following section the TD-MC-srPBE method is further tested against the above-mentioned functionals,
using the full test set in Figure \ref{molecules}.  

Considering the present selection of molecules, our initial study reveals promising results for the SOPPA-srPBE method. 
The method is a viable alternative to TD-MC-srPBE, 
showing often similar or even better accuracy, in particular for the DMABN and PP molecules. However, it should 
be noted that the molecules within the current test set are at large dominated by a single configuration 
and the present accuracy is not expected to extend to molecules exhibiting multiconfigurational character in their electronic 
ground state. 

\subsection{Performance of TD-MC-srPBE on the full molecular test set} \label{srMCSCFpeform}

\begin{figure}[htb!]
   \includegraphics[height=6cm]{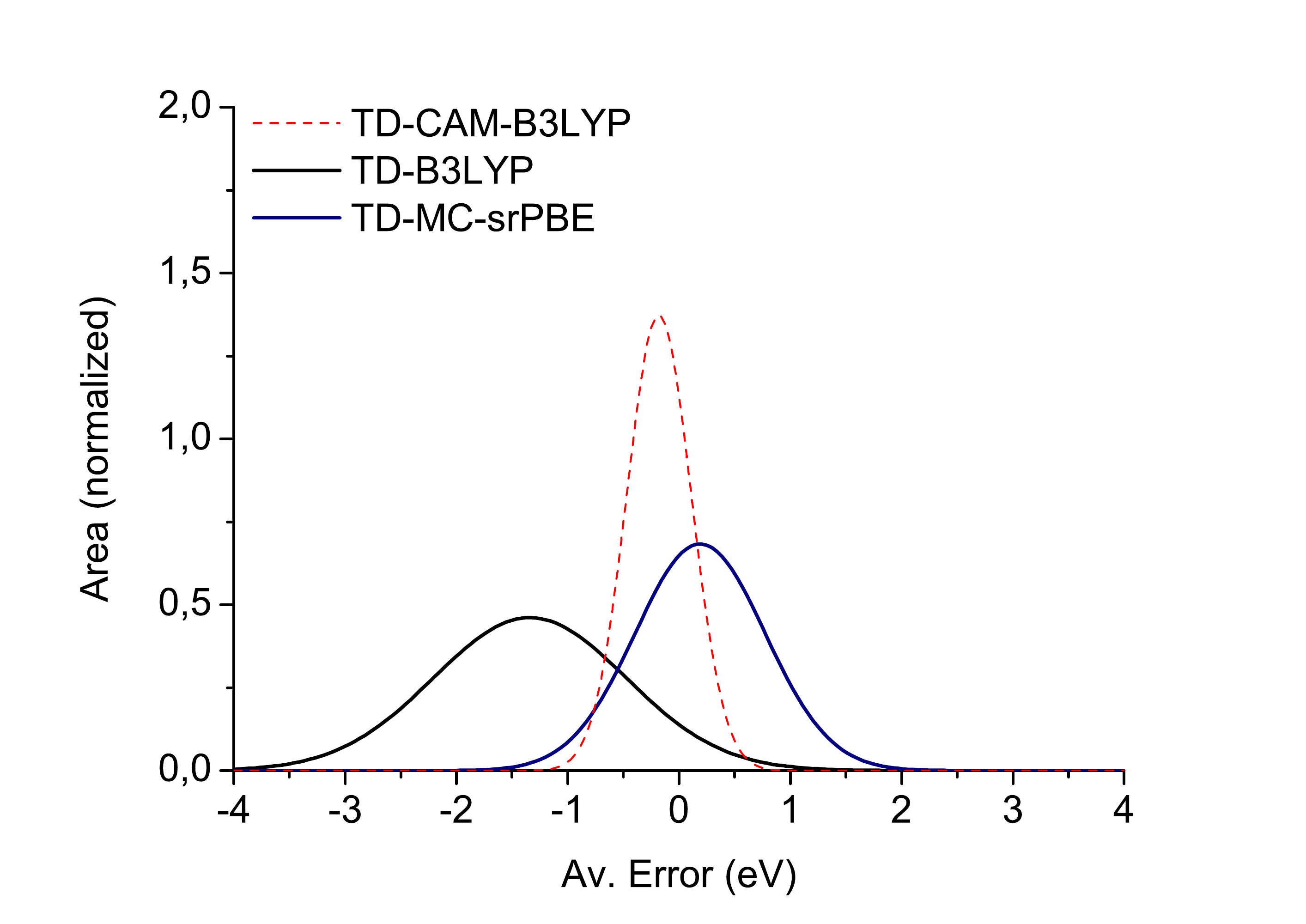}
 \caption{Normal distribution from data in Tables \ref{ErrorCT} (CT excitations) centered around the mean deviation.\label{GaussCTCombined}}
\end{figure}

\begin{figure}[htb!]
   \includegraphics[height=6cm]{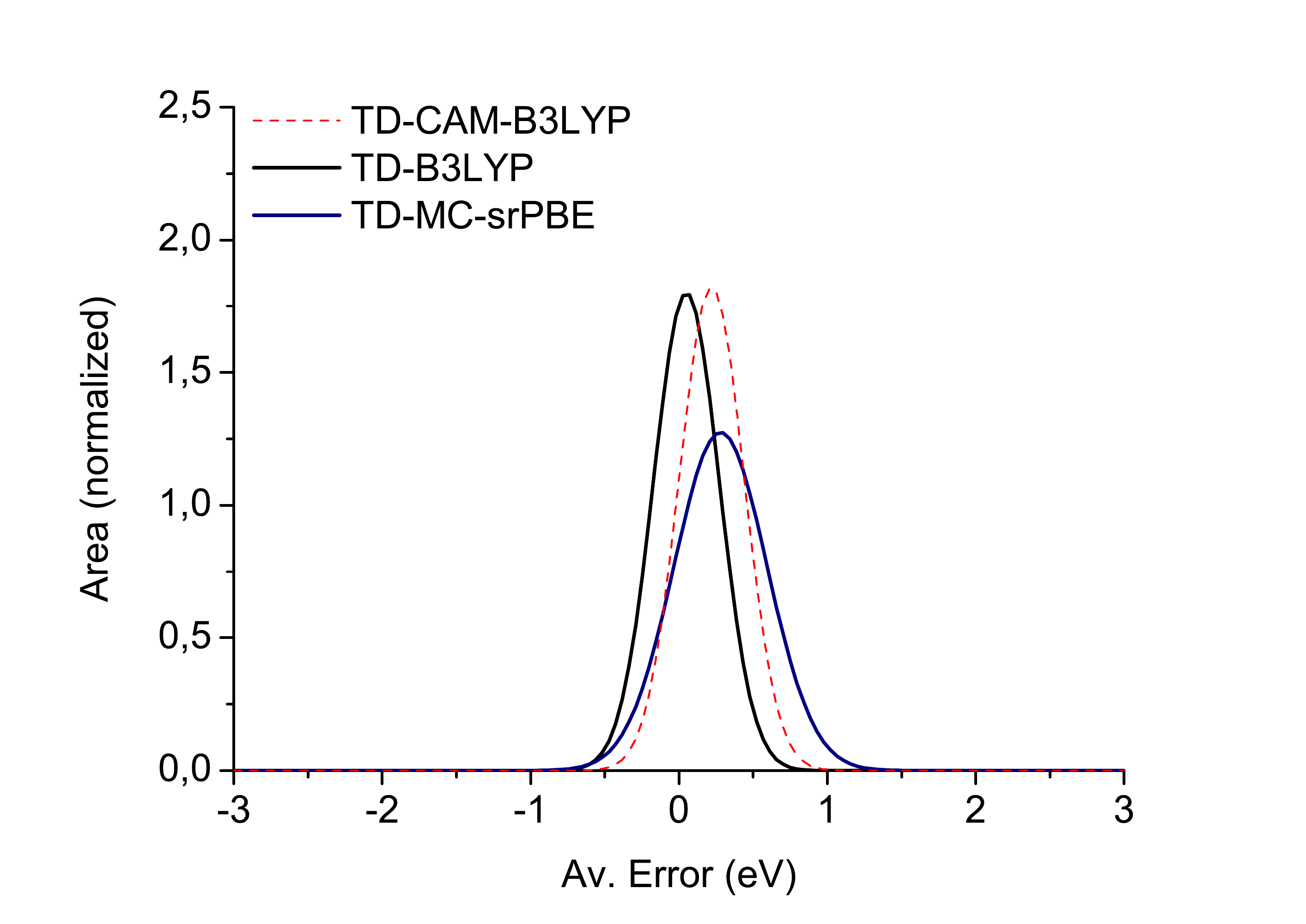}
 \caption{Normal distribution from data in Table \ref{ErrorLoc} (Local
 excitations) centered around the mean deviation.\label{GaussLocCombined}}
\end{figure}

The inclusion of the inorganic diatom \ce{HCl}, the $\beta$-dipeptide and the tripeptide for
testing the performance of the TD-MC-srPBE  method yields a total of 24 singlet excitations;
14 of these have charge transfer character and 10 are local. 
This test set is still not very extensive but we believe it is sufficiently large to compare 
TD-MC-srPBE with B3LYP and CAM-B3LYP performances on a reasonable statistical basis.
\begin{table}[htb!]
\caption{Error analysis for 24 excitations described in text. ``sr'' is short-hand for ``srPBE''. All errors are given in eV \label{Errortotal}}
 \begin{tabular}{llcccccccc}
    \hline \hline \\[-1.5ex]
             & TD-MC-sr & TD-B3LYP & TD-CAM-B3LYP  \\[0.5ex]
    \hline \\[-1.5ex] 
Mean         &   0.23    & -0.76 & -0.01  \\[0.5ex]
std. dev.    &   0.48    &  0.97 &  0.33  \\[0.5ex]
MAD          &   0.42    &  0.86 &  0.25  \\[0.5ex]
std. dev.    &   0.31    &  0.86 &  0.21  \\[0.5ex]
    \hline \hline
 \end{tabular}
\centering
\end{table}
\begin{figure}[htb!]
 \centering
 \includegraphics[height=6cm]{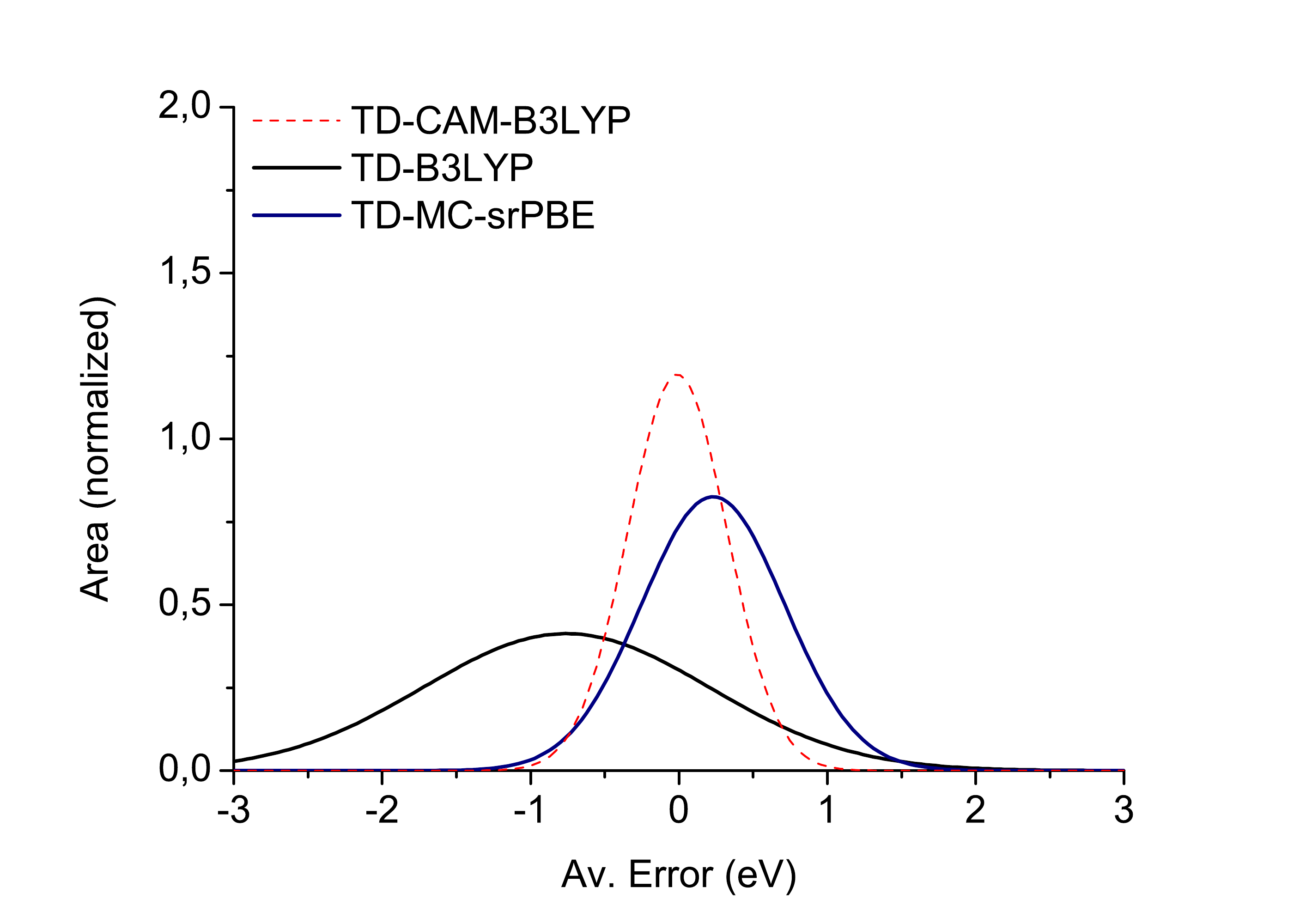} 
 \caption{Normal distribution from data in Table \ref{Errortotal}.\label{GaussTotCombined} centered around the mean deviation.}
\end{figure} 
For the full benchmark set, including both local and charge transfer
excitations, the statistical parameters are given in Table \ref{Errortotal} and 
the normal distributions are shown in Figure \ref{GaussTotCombined}. 
\begin{table}[htb!]
\caption{Error analysis for 14 CT excitations described in text. ``sr'' is short-hand for ``srPBE''. All errors are given in eV \label{ErrorCT}}
 \begin{tabular}{lccccccccc}
    \hline \hline \\[-1.5ex]
             & TD-MC-sr & TD-B3LYP & TD-CAM-B3LYP  \\[0.5ex]
    \hline \\[-1.5ex] 
Mean         &  0.19   & -1.34 & -0.18 \\[0.5ex]
std. dev.    &  0.58   &  0.86 & 0.29  \\[0.5ex]
MAD          &  0.51   &  1.36 & 0.27  \\[0.5ex]
std. dev.    &  0.31   &  0.85 & 0.22  \\[0.5ex]
    \hline \hline
 \end{tabular} 
\centering
\end{table}
The TD-MC-srPBE method generally shows good performance over the whole set, even with the moderate active spaces used here. 
In this aspect it is worthwhile to notice that 
the CASPT2 reference calculations for the peptide model systems used a significantly larger active space (although also a smaller basis set).   
The CAM-B3LYP functional is the most accurate with a very small mean deviation of -0.01 eV. 
Also the mean absolute deviation (MAD) is the smallest for CAM-B3LYP. 
B3LYP is significantly off as expected due to the charge transfer type of excitations. 
If focus is solely on these type of excitations, the error of B3LYP is even more pronounced, as shown from
the statistical analysis result over the charge transfer excitations in Table \ref{ErrorCT} (the normal distributions are displayed in Figure \ref{GaussCTCombined}). 
B3LYP now (on average) underestimates the vertical excitations by -1.34 eV, whereas CAM-B3LYP still underestimates CT type excitations, but with a considerable smaller margin.
TD-MC-srPBE is here comparable to CAM-B3LYP (although the MAD is
somewhat higher) and it seems that in general the srPBE functional remedies the commonly encountered 
overestimation of excitation energies at the MCSCF level.
For completion the results from the local excitations are given in Table
\ref{ErrorLoc} and Figure \ref{GaussLocCombined}.
\begin{table}[htb!]
\caption{Error analysis for 10 local excitations described in text. ``sr'' is short-hand for ``srPBE''.  
 All errors are given in eV \label{ErrorLoc}}
 \begin{tabular}{lccccccccc}
    \hline \hline \\[-1.5ex]
                & TD-MC-sr & TD-B3LYP & TD-CAM-B3LYP  \\[0.5ex]
    \hline \\[-1.5ex] 
Mean            &  0.28   & 0.05  & 0.22  \\[0.5ex]
std. dev.       &  0.31   & 0.22  & 0.22  \\[0.5ex]
MAD             &  0.30   & 0.16  & 0.22  \\[0.5ex]
std. dev.       &  0.29   & 0.15  & 0.21  \\[0.5ex]
    \hline \hline
 \end{tabular} 
\centering
\end{table}
B3LYP is here on average the closest to the reference data (the deviation is 0.05 eV). 
 It is noteworthy that TD-MC-srPBE is still accurate, although not 
as accurate as B3LYP and CAM-B3LYP. As it was the case for SOPPA-srPBE it should be noted that the use of B3LYP and 
CAM-B3LYP will be problematic for molecules showing significant multireference and/or double excitation character.

\subsection{The retinal chromophore}

\begin{figure*}[htb!]
\begin{minipage}[t]{.25\textwidth}
 \begin{tabular}{cr@{}ccc}
    \hline \hline \\[-1.5ex]
Config.    & Coeff.         & & & Assign.                       \\[0.5ex]
    \hline \\[-1.5ex]  
  \multicolumn{5}{c}{$S_{0}\rightarrow S_{1}$ }            \\[0.5ex]
 $\bm{1}$  &   0.778        & & & $\pi_{3}(1)\rightarrow\pi^{*}_{4}(1)$   \\[0.5ex]
 $\bm{2}$  &  $-$0.299      & & & $\pi_{2}(1)\rightarrow\pi^{*}_{4}(1)$                  \\[0.5ex]
 $\bm{3}$  &   0.247        & & & $\pi_{3}(2)\rightarrow\pi^{*}_{4}(2)$                  \\[0.5ex]
$\bm{4}$   &   0.230        & & & $\pi_{3}(1)\rightarrow\pi^{*}_{5}(1)$                  \\[0.5ex]
    \hline \\[-1.7ex]  
  \multicolumn{5}{c}{$S_{0}\rightarrow S_{2}$ }            \\[0.5ex]
$\bm{1}$   &   0.458        & & & $\pi_{3}(1)\rightarrow\pi^{*}_{4}(1)$                   \\[0.5ex]
$\bm{2}$   &  $-$0.299      & & & $\pi_{2}(1)\rightarrow\pi^{*}_{4}(1)$                  \\[0.5ex]
$\bm{3}$   &  $-$0.416      & & & $\pi_{3}(2)\rightarrow\pi^{*}_{4}(2)$                  \\[0.5ex]
$\bm{4}$   &  $-$0.147      & & & $\pi_{3}(1)\rightarrow\pi^{*}_{5}(1)$                  \\[0.5ex]
    \hline \hline
 \end{tabular}
\end{minipage}
\begin{minipage}[t]{.74\textwidth}
\vspace{-2.7cm}\includegraphics[height=5.5cm]{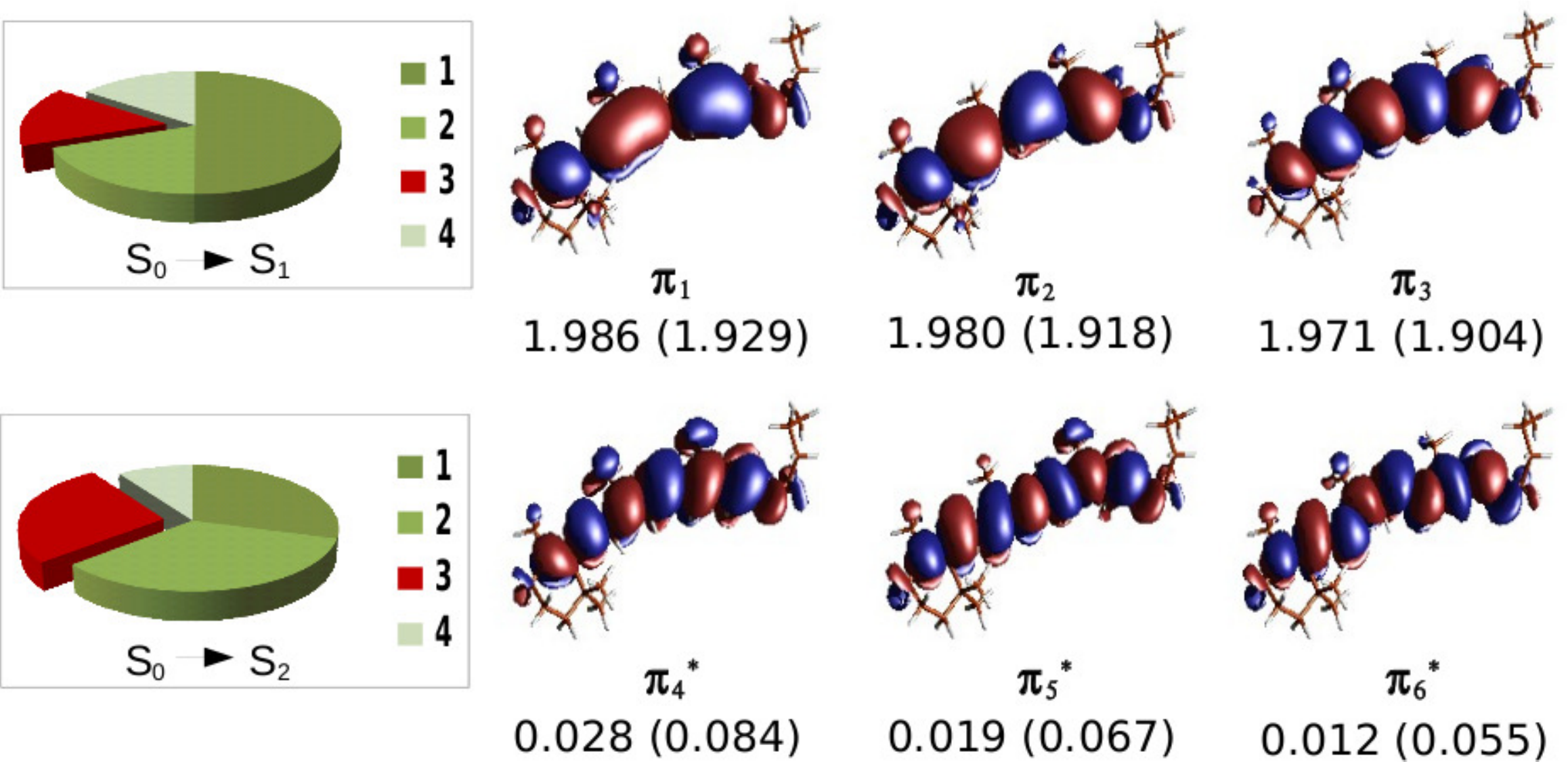}
\end{minipage}
\caption{Orbitals within the active space for the retinal chromophore in Figure \ref{retinal_chromophore}. Numbers under the
orbitals are the MP2-srPBE natural orbital occupancies. Regular MP2
occupancies are in parentheses. The table at the left-hand side shows linear response coefficients for the dominant configurations, \textbf{1}--\textbf{4}. The numbers
in parentheses are the number of involved electrons. The chart shows the relative contributions of  \textbf{1}--\textbf{4} for both the $S_{0}\rightarrow S_1$ and 
$S_{0}\rightarrow S_2$ excitations, where the red (\textbf{3}) is double excitation character.   
\label{retinal_nos}}
\end{figure*}

As a final test case, we applied the TD-MC-srPBE method
to the calculation of the low-lying singlet excited state spectrum of the 
retinal chromophore. This chromophore displays significant multireference character in its 
ground state, whereas the low-lying singlet excitations are dominated by a double
excitation character, which cannot be described by regular TD-DFT (within the common adiabatic approximation).   
The natural orbitals spanning the chosen CAS(6,6) space are shown in
Figure \ref{retinal_nos} and display the expected increase in nodal
planes as one moves from orbitals of 
high occupation numbers ($\pi_{1}$--$\pi_{3}$) towards orbitals of lower
occupation numbers ($\pi^{*}_{4}$--$\pi^{*}_{6}$). 
Before discussing the excitation energies in detail, a technical aspect concerning the choice of active space in TD-MC-srPBE is addressed. 
A well-known problem with including dynamical correlation on top of a multireference method (for example CASSCF/CASPT2) 
is that it can lead to intruder states or root flipping. In MC-srDFT
dynamical and static correlations are treated simultaneously,
which often means that an active space can be used that is significantly smaller than the
one of a regular MCSCF calculation. 
This beneficial feature is illustrated by the MP2-srPBE and MP2 (in parentheses) natural
orbital occupation numbers of the three highest occupied orbitals 
shown in Figure \ref{retinal_nos}. Similar differences between MP2 and MP2-srPBE 
have been observed for all molecules considered in this study, and a comparison of MP2 and MP2-srPBE occupation numbers 
is given in the supporting information for the full test set\cite{supporting} 

Our excitation energies for the retinal chromophore computed at the TD-MC-srPBE level are compiled in Table \ref{retinal_table}. 
As can be seen from Table \ref{retinal_table}\ the singlet excited states S$_{1}$\ and S$_{2}$\ are well separated and the first state
is the bright state with a large oscillator strength whereas the second state is the dark state with 
a considerably lower oscillator strength. Using the nomenclature from polyenes, the S$_1$\ state thus corresponds to the $B_{u}$ state while the second state S$_2$\ is the $A_{g}$ state, which is in 
agreement with both experiment\cite{birge1990,andersen2005,nielsen2006} and previous CASPT2 results\cite{cembran2005} using the same basis set.
Quantitatively, the S$_{0}\rightarrow$ S$_{1}$\ excitation is in good agreement with previous CASPT2 and other theoretical results 
(c.f.\ footnote $a$ in Table \ref{retinal_table}). We note that the experimental value given here is the gas-phase value, while we have used a 
geometry obtained in an optimization considering also the surrounding protein (the calculation itself does \textit{not} include the protein environment) taken from a forthcoming publication. 
Thus one should not expect a one-to one correspondence which should be kept in mind when considering 
 the $S_{0}\rightarrow$ S$_{2}$\ excitation energy. 
 The latter is slightly overestimated by 0.41 eV at the TD-MC-srPBE level compared to the 
 experimental gas-phase value but also to the theoretical value of Altun and co-workers\cite{altun2008}). 
However, the agreement with theory must be still considered
reasonable in light of employing slightly different retinal models, quantum mechanical methods as well as geometry optimization conditions. 

\begin{table}[htb!]
\caption{Excitation energies for the retinal chromophore (eV) with oscillator strengths in parentheses. ``sr'' is short-hand for ``srPBE''. \label{retinal_table}}
 \begin{tabular}{llcccccccc}
    \hline \hline \\[-1.5ex]
 Excitation                  & TD-MC-sr           &  DDCI2 + Q$^{a}$          & Exp.  \\[0.5ex]
    \hline \\[-1.5ex]  
$S_{0}\rightarrow S_{1}$  &  2.29 (1.597)   & 2.27\cite{altun2008}          & 2.03\cite{andersen2005} \\[0.5ex]
$S_{0}\rightarrow S_{2}$  &  3.63 (0.522)   & 3.07\cite{altun2008}          & 3.22\cite{nielsen2006}  \\[0.5ex]
    \hline \hline
 \end{tabular} 
\begin{tablenotes}\footnotesize 
\item[a]$^{a}$The DDCI2 calculations were performed with an underlying CAS(12,12). For the for $S_{0}\rightarrow S_{1}$ excitation CASPT2(12,12) obtains 2.32 eV\cite{cembran2005} 
while B3LYP obtains 2.48\cite{vreven2003} eV. 
 \end{tablenotes}
\end{table}
The retinal calculations nicely illustrate an important aspect of the TD-MC-srPBE method. 
Both the $S_{0}\rightarrow S_{1}$ and the $S_{0}\rightarrow S_{2}$ excitations 
have a considerable doubly-excited character, as indicated by the
significant
weight of configuration $\bm{3}$ (red) in the charts of Figure
\ref{retinal_nos} (for the linear response coefficient of this configuration, see the
accompanying table on the left-hand side of Figure \ref{retinal_nos}).
Indeed, this weight is so important in the 
$S_{0}\rightarrow S_{2}$ transition that the latter can be considered 
as a two-electron $\pi_{3}\rightarrow\pi^{*}_{4}$ excitation. 
TD-DFT based on its standard adiabatic approximation formulation 
cannot describe such a transition, ultimately missing the electronic nature of the dark state.  
It should be noted that the present study lacks the effect from the
protein environment which can be significant as studies by S{\"o}derhjelm \textit{et al.} have shown\cite{soderhjelm2009}. 
Work to incorporate the effect from the environment into our TD-MC-srDFT
model is currently in progress based on the polarizable-embedding method by Kongsted and co-workers\cite{olsen2010,olsen2011}.

\section{Conclusion}\label{conclusion}

In this paper the SOPPA-srDFT method has been formulated and tested
together with the recently presented TD-MC-srDFT approach using a srPBE functional for the srDFT part.
We have compared the performance of these methods to standard TD-DFT using B3LYP and CAM-B3LYP functionals for excitation
energies, using a model peptide, $N$-phenyl pyrrole (PP) and 4-($N$,$N$-dimethylamino) benzonitrile (DMABN) as test cases. 
The assessment has been done with explicit focus on charge-transfer 
excitations although results for local excitations have been included as well. 
While the regular SOPPA method underestimates both local and charge-transfer excitations, 
SOPPA-srPBE is generally much closer to the reference CASPT2 data. 
Considering the total benchmark set of 24 excitations (from molecules in Figure \ref{molecules}) TD-DFT/CAM-B3LYP still performs best 
whereas due to the large discrepancies in the charge-transfer excitations, TD-DFT/B3LYP 
cannot be recommended for a general application to excitation energies of various characters. 
The TD-MC-srPBE method commonly yields sufficiently accurate charge-transfer excitation energies 
while in some cases it even outperforms TD-DFT/CAM-B3LYP. 
Notably, this accuracy can not only be achieved with quite small
active spaces for the long-range-interacting CASSCF wave function but the  
MC-srDFT \textit{ansatz} also scales nearly with respect to system size compared to regular MCSCF. 

Doubly-excited (singlet) states cannot be described with regular TD-DFT schemes if they
rely on the popular adiabatic approximation. The TD-MC-srPBE method on the other hand does not suffer from this 
shortcoming by design since double excitation can be effectively described 
within the long-range MCSCF part of the wave function. In order to illustrate this important capability, 
we have here investigated the retinal chromophore as a prime example of (bio-)chemical interest 
where double excitations play a major role in the photophysics of the low-lying excited states.  
Our present results for the excitation energies of the first two singlet excited states are promising 
and within the range of previously reported CASPT2 and MRCI data, 
albeit the fact that the latter methods required much larger active spaces. 
To further enhance the scope of TD-MC-srDFT applications we currently address the computation of properties which
are not easily implemented for CASPT2 type wave functions (such as NMR parameters). An extension of the 
TD-MC-srDFT approach to embedding into solvent or protein environments is in progress in our laboratories. 

\begin{acknowledgments}

E.D.H. thanks OTICON and Augustines funds for stipends.
The authors wish to thank the Danish Center for Scientific Computing for
computational resources. S.K.~acknowledges the Danish Natural Science Research Council for an individual postdoctoral grant 
(10-082944). E.F. thanks ANR (DYQUMA project). 

\end{acknowledgments}



\newcommand{\Aa}[0]{Aa}

\end{document}